\documentclass[reprint, aps,prc,twocolumn,showpacs,showkeys, 10pt, longbibliography, nolinenumbers, floatfix]{revtex4-1}

\usepackage{amsmath}
\usepackage{graphicx}
\usepackage{float}
\usepackage{dcolumn}
\usepackage{multirow}
\usepackage{natbib}
\usepackage[colorlinks = true,linkcolor = blue,urlcolor  = blue,citecolor = blue,anchorcolor = blue]{hyperref}
\usepackage{enumerate}
\begin{document}

\title{Speed of sound constraints from tidal deformability of neutron stars}

\author{A. Kanakis-Pegios}
\email{alkanaki@auth.gr}

\author{P.S. Koliogiannis}
\email{pkoliogi@physics.auth.gr}

\author{Ch.C. Moustakidis}
\email{moustaki@auth.gr}

\affiliation{Department of Theoretical Physics, Aristotle University of Thessaloniki, 54124 Thessaloniki, Greece}

%%%%%%%%%%%%%%%%%%%%%%%%%%%%%%%%%%%%%%%%%%%%%%%%%%%%%%%%%%%%%%%%%%%%%%%%%%%%%%%%
\begin{abstract}
%%%%%%%%%%%%%%%%%%%%%%%%%%%%%%%%%%%%%%%%%%%%%%%%%%%%%%%%%%%%%%%%%%%%%%%%%%%%%%%%	
The upper bound of the speed of sound in dense nuclear matter is one of the most interesting but still  unsolved problems in Nuclear Physics. 
Theoretical studies in connection with recent observational data 
of isolated neutron stars as well as binary neutron stars systems offer an excellent opportunity to shed light on this problem. 
In the present work, we suggest a method to directly relate the measured tidal deformability (polarizability) of binary neutron stars system (before merger) to the  maximum neutron star  mass scenario and  possible upper bound on the speed of sound. 
This method is based on the simple but efficient  idea that while the upper limit of the effective tidal deformability favors soft equations of state, the recent high measured  values of neutron star mass favor stiff ones. In the present work, firstly, using a simple well established   model we parametrize  the stiffness of the equation of state
with the help of the speed of sound. Secondly, in comparison with the recent observations by LIGO/VIRGO collaboration of two  events,  
GW170817   and GW190425,  we suggest possible robust constraints. Moreover, we evaluate and postulate, 
in the framework of the present method,   what kind of future measurements  could  help us to improve the stringent of the constraints on the neutron star equation of state.   

\pacs{26.60.-c, 21.65.+f, 04.30.Tv, 97.60.Jd}

\keywords{Equation of state; Speed of sound; Neutron stars; Tidal deformability}
\end{abstract}

\maketitle

%%%%%%%%%%%%%%%%%%%%%%%%%%%%%%%%%%%%%%%%%%%%%%%%%%%%%%%%%%%%%%%%%%%%%%%%%%%%%%%%

%%%%%%%%%%%%%%%%%%%%%%%%%%
\section{Introduction}
%%%%%%%%%%%%%%%%%%%%%%%%%%%
The properties of dense nuclear matter still remain one of the unsolved problems in Nuclear Physics.
Due to the limitation of terrestrial experiments, concerning the  high density behavior of nuclear matter, the scientific community focuses  on studying compacts objects of the Universe, including mainly white dwarfs and neutron stars. In particular, neutron stars considered as the best extraterrestrial laboratories to study unexplored properties of dense matter~\cite{Shapiro-1983,Glendenning,Haensel,Weinberg}. One of these properties is the limit of the speed of sound in dense matter and in particular the exploration of possible upper bounds predicted  by various theoretical models or even intuition  conjectures. 

To be more specific, the main assumption is that the speed of sound  cannot exceed the speed of light
because of  causality. However, this is not a rigorous proof.
Zel'dovich~\cite{Zeldovich-62,Zeldovich-71} was the first pointed out the importance to define a rigorous speed of sound limit upon the equation of state (EoS). Firstly, he   postulated  that in the case of electromagnetic interaction, the covariant formulation suggests that the assumption $v_s \leq c/\sqrt{3}$ is a general low of nature. Then, using some simple assumptions, he stated  that considering the case of interaction of the baryons through  a vector field, the upper bound of the speed of sound is the speed of light, $v_s=c$. His main conclusion was that  in the domain, where our knowledge is limited, the only restriction imposed by general principles is that $v_s\leq c$~\cite{Zeldovich-62,Zeldovich-71}. Hartle~\cite{Hartle-1978}, in an interesting and extensive study, pointed out that causality is not enough to constrain the high-density part of the EoS. Weinberg \cite{Weinberg} showed that the speed of sound is much less than the speed of light for a cold nonrelativistic fluid. Although it increases with temperature, it does not exceed the value $v_s=c/\sqrt{3}$ at the limit of very high temperatures. Recently, Bedaque and Steiner~\cite{Bedaque-2015} have provided simple arguments that support the limit $v_s=c/\sqrt{3}$    in nonrelativistic and/or weakly coupled theories. This was demonstrated in several classes of strongly coupled theories with gravity duals. The upper limit saturated only in conformal theories. In particular, in Ref.~\cite{Bedaque-2015} the authors found also that  the existence of neutron stars with masses about two solar masses, combined with the knowledge of the EoS of hadronic matter at low densities, is not consistent with the bound $c/\sqrt{3}$.
The effects and possible constraints on the speed of sound on the tidal deformability have been analyzed and discussed also  in Refs.~\cite{Moustakidis-2017,Reed-2020,Oeveren-2017,Ma-2019}. \\
\indent Summarizing, there are two controversial considerations for the upper bound in the speed of sound of dense matter. In first one, the only limitation is imposed by the demand of causality where the speed of sound should not exceed the one of light.  Other non relativistic models, employing microscopic interactions lead also to uncasual  EoSs (the most remarkable example is the EoS of Akmal {\it et al.}~\cite{Akmal-98}). In the second case, various theoretical considerations predict as  the upper limit on the speed of sound the value $v_s=c/\sqrt{3}$. 
There is also an additional, important reason to look out for constraints on the speed of sound and especially for the lower limit $v_s=c/\sqrt{3}$, since is related to the existence or not of quark-matter cores in neutron stars or, in general, to the existence of quark stars (for a review see Refs.~\cite{Baym-2018,Annala-20}).
Finally, is worth mentioning that there are other theories where the possibility that the speed of sound is exceeding that of light in ultradense matter, is examined (see for example Ref.~\cite{Bludman-68}). However, in the present study we are not going to discuss the implications of this interesting possibility.  

The main motivation of the present work is the implementation of a method directly related to the measured tidal deformability (polarizability) of binary neutron stars system (before merger), to the  maximum neutron star mass  measurements  and  also to possible upper bounds on the speed of sound. The method is based on the simple but efficient  idea that while the upper limit of the effective tidal deformability favors soft EoSs, the recent  high measured values of neutron star mass favor stiff ones. The starting  point is the use of a  simple model where the neutron star EoS  parametrized with the help of various speed of sound bounds, and in this way its stiffness (softness), is a functional of the speed of sound and the transition density. The predictions are combined with the recent observations of two  events,  GW170817   and GW190425~\cite{Abbott-1,Abbott-2,Abbott-3,Abbott-4}, as well as the current observed maximum neutron star masses ($1.908\pm 0.016\;M_{\odot}$~\cite{Arzoumanian-2018}, $2.01\pm 0.04\;M_{\odot}$~\cite{Antoniadis-2013}, $2.14^{+0.10}_{-0.09}\;M_{\odot}$~\cite{Cromartie-2019}, $2.27^{+0.17}_{-0.15}\;M_{\odot}$~\cite{Linares-2018}) (for more details see Ref.~\cite{Koliogiannis-2019}). 
The interplay between the demand of a) a soft EoS for low densities (in order to be in accordance with the upper limit of the tidal deformability) and b) a stiff EoS for high densities (in order to predict the high neutron star measurements) leads to robust constraints on the EoS. Moreover, employing the present method we will be able to  postulate what kind of future measurements  could  help us to improve the stringent of the constraints.

In view of the previous statement, we mention the very recent observation of the GW190814 event,where a gravitational wave has been detected from the coalescence of a $22.2-24.3\;M_{\odot}$ black hole with a non-identified compact object with  mass $2.5-2.67\;M_{\odot}$~\cite{Abbott-5,Tan-2020}. Although the authors suggest that is unlikely the secondary mass to belong to a neutron star,  they do leave {\it open the window}  that the improved knowledge of the neutron star EoS and further observations  of the astrophysical population of compact objects could alter this assessment.

The paper is organized as follows: In Sec. II we briefly present the model for the parametrization of the EoS with the help of the speed of sound which leads also to the  maximum mass configuration. In Sec. III we present the basic formalism related to the tidal deformability  used in the present study.  The results are presented and discussed in Sec. IV. The main conclusions of the present study are presented in Sec. V.
  
%%%%%%%%%%%%%%%%%%%%%%%%%%%%%%%%%%%%%%%%%%
\section{Speed of sound bounds and neutron star EoS}
%%%%%%%%%%%%%%%%%%%%%%%%%%%%%%%%%%%%%%%%%

We have constructed the maximum mass configuration   with the properly parametrization of the  neutron star EoS~\cite{Margaritis-2020,Rhoades-1974,Kalogera-1996,Koranda-1977,Chamel-2013a,Breu-2016,Alsing-2018,Podkowka-2018,Xia-2019}
\begin{eqnarray}
P({\cal E})&=&\left\{
\begin{array}{ll}
P_{\rm crust}({\cal E}), \quad  {\cal E} \leq {\cal E}_{\rm c-edge}  & \\
\\
P_{\rm NM}({\cal E}), \quad  {\cal E}_{\rm c-edge} \leq {\cal E} \leq {\cal E}_{\rm tr}  & \\
\\
\left(\frac{v_{\rm s}}{c}  \right)^2\left({\cal E}-{\cal E}_{\rm tr}  \right)+
P_{\rm NM}({\cal E}_{\rm tr}), \quad  {\cal E}_{\rm tr} \leq {\cal E}  . & \
\end{array}
\right.
\label{EOS-1}
\end{eqnarray}
where $P$ and ${\cal E}$ are the pressure and energy density, respectively, and $\mathcal{E}_{\rm tr}$ is the transition energy density. In region ${\cal E} \leq {\cal E}_{\rm c-edge}$, we used the equation of Feynman {\it et al.}~\cite{Feynman-1949} and also of Baym {\it et al.}~\cite{Baym-1971} for the crust and low densities of neutron star. In the  intermediate region, ${\cal E}_{\rm c-edge} \leq {\cal E} \leq {\cal E}_{\rm tr}$, we employed a specific EoS based on the MDI model and data from Akmal {\it et al.}~\cite{Akmal-98}, while for ${\cal E}_{\rm tr}\geq \mathcal{E}$, the EoS is maximally stiff with the speed of sound, defined as  $v_s=c\sqrt{\left(\partial P / \partial {\cal E}\right)}_{\rm S}$ (where $S$ is the entropy)  fixed in the present work on the two values, $c/\sqrt{3}$ and $c$. Obviously, the implementation of speed of sound values between these two limits will lead to results well constrained by the two mentioned limits. Although the energy densities below the ${\cal E}_{\rm c-edge}$ have negligible effects on the maximum mass configuration, we used them in calculations for the accurate estimation of the tidal deformability. The cases which took effect in this study were the ones where the fiducial baryon transition density is $n_{\rm tr} = p n_{\rm 0}$, where $n_{0}$ is the saturation density of symmetric nuclear matter ($n_{\rm 0}= 0.16$ fm$^{-3}$) and $p$ takes the values $1,1.5,2,3$. The predicted EoSs are functional of $n_{\rm tr}$ and $v_s$ and implemented to study their effects on the bulk neutron star properties including $M_{\rm max}$ (maximum mass of a neutron star), $R_{\rm max}$ (its corresponding radius),  $R_{1.4}$ (radius of a $1.4\;M_\odot$ neutron star), $\Lambda$ (tidal deformability; for definition see below Eq.~(\ref{Lamb-1})) etc.

In approach (\ref{EOS-1}) the continuity on the  EoS is well ensured but due  to its artificial character,  the continuity in the speed of sound at the transition density is not. It is worth to point out that, since the speed of sound is involved in the calculation of the tidal deformability,  must be treated very carefully especially in regions of discontinuity~\cite{Postnikov-2010}.  Therefore, in order to ensure the continuity and  a smooth phase transition, we employ a method presented in Ref.~\cite{Tews-2018}. We proceeded with the matching of the EoSs on the transition density  by considering that, above this value, the speed of sound is parametrized as follows (for more details see Ref.~\cite{Tews-2018})
\begin{equation}
\frac{v_{\rm s}}{c}=\left(a-c_1\exp\left[-\frac{(n-c_2)^2}{w^2} \right]\right)^{1/2}, \quad a=1, 1/3
\label{speed-matc-1}
\end{equation}
where the parameters $c_1$, $c_2$, and  $w$ are fit to the speed of sound and its derivative at $n_{\rm tr}$, and also to the demands $v_{\rm s}(n_{\rm tr})=[c, c/\sqrt{3}]$~\cite{Margaritis-2020}. Using Eq.~(\ref{speed-matc-1}), the EoS for $n \geq n_{{\rm tr}}$ can be constructed with the help of the following recipe~\cite{Tews-2018}
\begin{equation}
{\cal E}_{i+1} = {\cal E}_i+\Delta {\cal E}, \quad P_{i+1} = P_i+\left(\frac{v_s}{c}(n_i)\right)^2\Delta {\cal E},
\label{eq:5}
\end{equation}
\begin{equation}
\Delta {\cal E} = \Delta n\left(\frac{{\cal E}_i+P_i}{n_i} \right),
\label{eq:6}
\end{equation} 
\begin{equation}
\Delta n = n_{i+1}-n_i.
\label{eq:7}
\end{equation}
It is worth to mention that the results for the bulk neutron star properties of the approach where discontinuity  is presented and the one where continuity exhibits are presented in Table V of Ref.~\cite{Margaritis-2020}. The main conclusion was that the two approaches  converge and consequently the effects of the discontinuity are negligible.

%%%%%%%%%%%%%%%%%%%%%%%%%%%%%%
\section{Tidal deformability}
%%%%%%%%%%%%%%%%%%%%%%%%%%%%%%%
It is expected that one of the most important sources for ground-based gravitational wave detectors are the gravitational waves from the final stages of inspiraling binary neutron stars~\cite{Postnikov-2010,Baiotti-2019,Flanagan-08,Hinderer-08,Damour-09,Hinderer-10,Fattoyev-13,Lackey-015,Takatsy-2020}. The masses of the components of the system can be determined with moderate accuracy, especially if  neutron stars are slowly spinning during the early stages of the evolution. In particular,  Flanagan and Hinderer \cite{Flanagan-08} have  pointed out
that tidal effects are also potentially measurable during the early part of the evolution
when the waveform is relatively clean. 

The tidal fields induce quadrupole moments on neutron stars.
The response of the neutron star is described by the dimensionless so-called Love number $k_2$,
which depends on the neutron star structure and consequently on the mass and the EoS of the nuclear matter.
The tidal Love number $k_2$ is obtained from the ratio of the induced quadrupole moment $Q_{ij}$ to the applied tidal field $E_{ij}$~\cite{Flanagan-08,Thorne-1998}
\begin{equation}
Q_{ij}=-\frac{2}{3}k_2\frac{R^5}{G}E_{ij}\equiv- \lambda E_{ij},
\label{Love-1}
\end{equation}
where $R$ is the neutron star radius  and $\lambda=2R^5k_2/3G$ is the tidal deformability. The tidal Love number $k_2$ is given by \cite{Flanagan-08,Hinderer-08}
\begin{eqnarray}
k_2&=&\frac{8\beta^5}{5}\left(1-2\beta\right)^2\left[2-y_R+(y_R-1)2\beta \right] \nonumber \\
&\times&
\left[\frac{}{} 2\beta \left(6  -3y_R+3\beta (5y_R-8)\right) \right. \nonumber \\
&+& 4\beta^3 \left.  \left(13-11y_R+\beta(3y_R-2)+2\beta^2(1+y_R)\right)\frac{}{} \right.\nonumber \\
&+& \left. 3\left(1-2\beta \right)^2\left[2-y_R+2\beta(y_R-1)\right] {\rm ln}\left(1-2\beta\right)\right]^{-1}
\label{k2-def}
\end{eqnarray}
where $\beta=GM/Rc^2$ is the compactness parameter. The quantity $y_R$ is determined by solving the following differential equation
\begin{equation}
r\frac{dy(r)}{dr}+y^2(r)+y(r)F(r)+r^2Q(r)=0, 
\label{D-y-1}
\end{equation}
with the initial condition $ y(0)=2$~\cite{Hinderer-10}. $F(r)$ and $Q(r)$ are functionals of ${\cal E}(r)$, $P(r)$ and $M(r)$  defined as~\cite{Postnikov-2010,Hinderer-10}
\begin{equation}
F(r)=\left[ 1- \frac{4\pi r^2 G}{c^4}\left({\cal E} (r)-P(r) \right)\right]e^{2\mathcal{\lambda}(r)},
\label{Fr-1}
\end{equation}
and
\begin{eqnarray}
r^2Q(r)&=&\frac{4\pi r^2 G}{c^4} \left[5{\cal E} (r)+9P(r)+\frac{{\cal E} (r)+P(r)}{\partial P(r)/\partial{\cal E} (r)}\right]
\nonumber \\
&\times&e^{2\mathcal{\lambda}(r)}- 6e^{2\mathcal{\lambda}(r)} \nonumber \\
&-&\frac{4M^2(r)G^2}{r^2c^4}\left(1+\frac{4\pi r^3 P(r)}{M(r)c^2}   \right)^2e^{4\lambda(r)},
\label{Qr-1}
\end{eqnarray}
%%%%%
where
\begin{equation}
 e^{2\mathcal{\lambda}(r)}=\left(1-\frac{2M(r)G}{rc^2}  \right)^{-1}, 
\label{metricfun}
\end{equation}
is the metric function for a spherical star~\cite{Glendenning}.

Eq.~(\ref{D-y-1})  must be integrated self consistently  with the Tolman-Oppenheimer-Volkov (TOV) equations using the boundary conditions $y(0)=2$, $P(0)=P_c$ and $M(0)=0$~\cite{Postnikov-2010,Hinderer-08}. The solution of the TOV equations provides the mass $M$ and radius $R$ of the neutron star, while the corresponding solution of the differential Eq.~(\ref{D-y-1}) provides the value of $y_R=y(R)$. The latter along with the quantity $\beta$ are the  basic ingredients  of the tidal Love number $k_2$.

One  of the binary parameters that is well constrained by the gravitational wave detectors is the chirp mass {\it $\mathcal{M}_c$}, which is a combination of the component masses~\cite{Abbott-1,Abbott-3}
\begin{equation}
\mathcal{M}_c=\frac{(m_1m_2)^{3/5}}{(m_1+m_2)^{1/5}}=m_1\frac{q^{3/5}}{(1+q)^{1/5}},
\label{chirpmass}
\end{equation}
where $m_1$ is the mass of the heavier component star and $m_2$ is the lighter's one. Hence, the binary mass ratio $q=m_2/m_1$ is within $0\leq q\leq1$.

The gravitational waves transfer the information about the tidal effects which in a binary system is characterized by the effective tidal deformability~\cite{Abbott-1,Abbott-3}
\begin{equation}
\tilde{\Lambda}=\frac{16}{13}\frac{(12q+1)\Lambda_1+(12+q)q^4\Lambda_2}{(1+q)^5},
\label{L-tild-1}
\end{equation}
%%%%%%%%%%%%%%%%%%
where the key quantity $q$ characterizes the mass asymmetry. Moreover, $\Lambda_i$ is the dimensionless deformability defined as~\cite{Abbott-1,Abbott-3}
\begin{equation}
\Lambda_i=\frac{2}{3}k_2\left(\frac{R_i c^2}{M_i G}  \right)^5\equiv\frac{2}{3}k_2 \beta_i^{-5}  , \quad i=1,2.
\label{Lamb-1}
\end{equation}
Now, by replacing in Eq.~(\ref{Lamb-1}) the value of $k_2$ from Eq.~(\ref{k2-def}), we found that  $\Lambda_i$  depends both on the compactness of the star as well as on the value of $y(R)$. 

It is worth to point out that $\Lambda_i$  depends directly on the stiffness of the EoS through the compactness $\beta$ and also indirectly through the speed of sound which appears in Eq.~(\ref{Qr-1}). Moreover, due to the dependence of $\Lambda_i$ on the values of $y(R)$ (which affected on the structure of the crust),  useful information can be gained for the observational estimation (or constraints) on the tidal deformability. In other words, the specific structure of an EoS defines not only the tread of the M-R diagrams but also the tread of $\Lambda-R$ and $\Lambda-M$ diagrams. To be more specific, the present study bases its inspiration on the combination of  constraints from (a) the very recent observations of the maximum mass of neutron stars and (b) tidal deformability derived by the LIGO observations,  in order to impose, if it is possible, robust constraints on the speed of sound in dense matter.

%%%%%%%%%%%%%%%%%%%%%%%%%%%%%%
\section{Results and Discussion}
\begin{figure}
\centering
\includegraphics[width=0.49\textwidth]{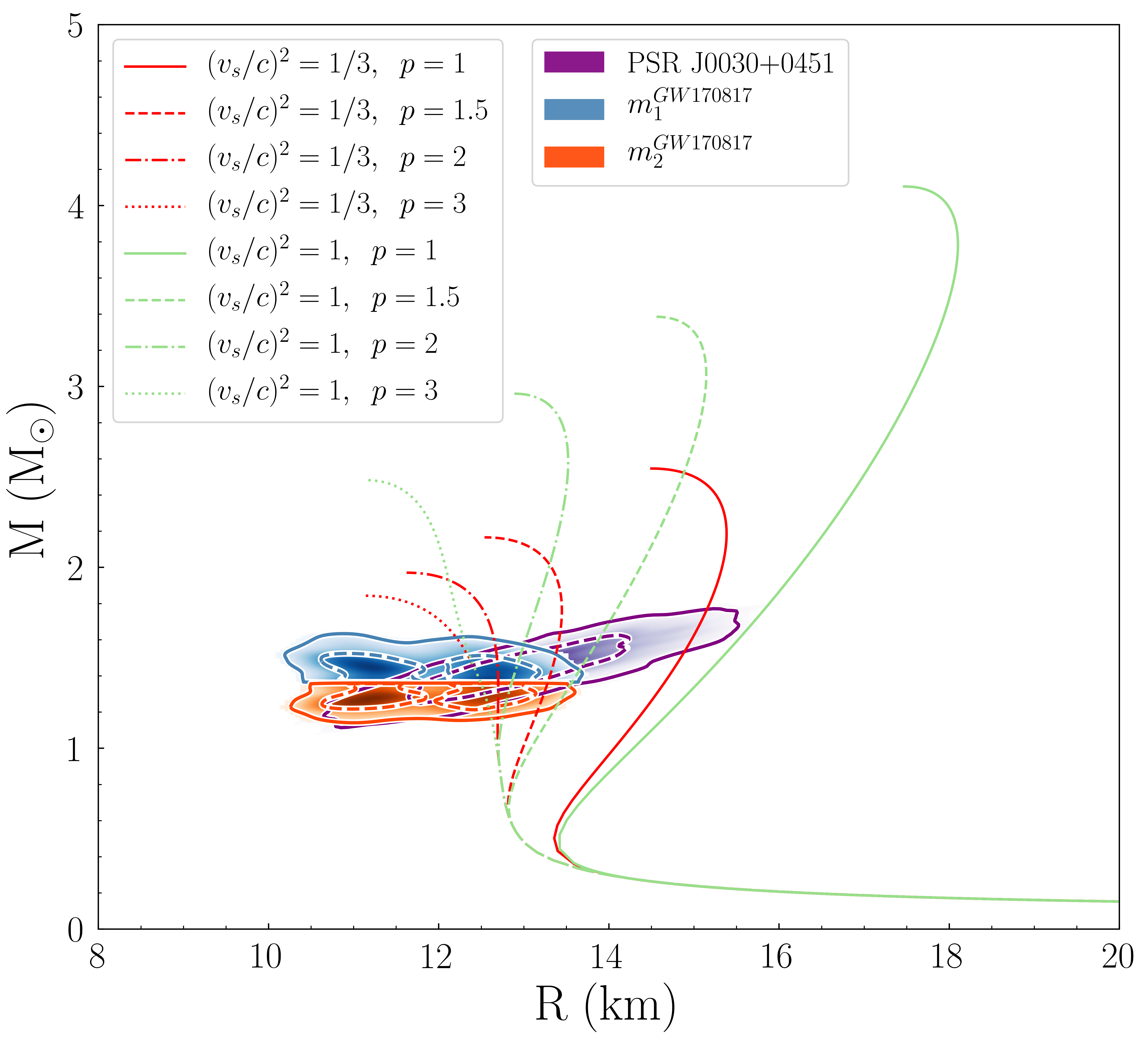}
\caption{Mass vs radius for an isolated neutron star, for the two cases of speed of sound. The green (red) lines correspond to the upper (lower) bound. The purple diagonal shaded region corresponds to NICER's observation (data taken from Ref.~\cite{Miller-2019}), while the blue upper (orange lower) shaded region corresponds to the higher (smaller) component of GW170817 event (data retrieved from Ref.~\cite{Abbott-2}). The solid (dashed) contour lines describe the $90\%$ ($50\%$) confidence interval.}
\label{MR-1}
\end{figure}

%%%%%%%%%%%%%%%%%%%%%%%%%%%%%%%
In our study we used two extreme scenarios for the value of speed of sound, the lower bound of $(v_s/c)^2=1/3$ and the upper one of $(v_s/c)^2=1$, and four transition densities $n_{{\rm tr}}=\{1,1.5,2,3\} n_0$~\cite{Margaritis-2020}. By solving numerically the system of TOV equations of hydrostatic equilibrium for an isolated cold neutron star, combined with the bounds above, we obtained the mass-radius diagram (see Fig.~\ref{MR-1}). In Fig.~\ref{MR-1} the red colored lines correspond to the $(v_s/c)^2=1/3$ limit, while the green ones correspond to the $(v_s/c)^2=1$ limit. The transition densities lead to bifurcations in M-R diagram. Between the same kind of linestyle, the upper (lower) bound $v_s=c$ ($v_c=c/\sqrt{3}$) of speed of sound corresponds to higher (lower)  masses.  As shown in Fig.~\ref{MR-1}, the higher the transition density, the softer the EoS, with the lower limit of $(v_s/c)^2=1/3$ leading to a more soft EoS, compare to the $(v_s/c)^2=1$ branch. In addition, the estimation of the GW170817 event and the NICER's data are also displayed~\cite{Abbott-2,Miller-2019}. One can observe that the GW170817 event of a binary neutron star merger provides stringent constraints, compared to data of NICER's observation. In particular, while  NICER favors stiffer EoSs,  GW170817 event favors softer. We notice that despite the fact that there is a significant overlap between the two observations, meaning that there is an agreement, the specific gravitational-wave's origin (GW170817) information is more suitable for our study. This is due to the fact that while the NICER's contour estimation covers almost all the cases, the neutron stars merger that was detected (GW170817), restricts the cases, excluding at least those with transition density $n_{{\rm tr}}=n_0$, for both cases of speed of sound. 

%%%%%%%%%%%%%%%%%%%%%%%%%%%%%%%%%%%%%%%%%%%%%%%%%%%%%%%%%
%FIGURE-1
%%%%%%%%%%%%%%%%%%%%%%%%%%%%%%%%%%%%%%%%%%%%%%%%%%%%%%%%%%%%%%%%%%%%%%
%%%%%%%%%%%%%%%%%%%%%%%%%%%%%%%%%%%%%%%%%%%%%%%%%%%%%%%%%
%FIGURE-2
\begin{figure}
\centering
\includegraphics[width=0.49\textwidth]{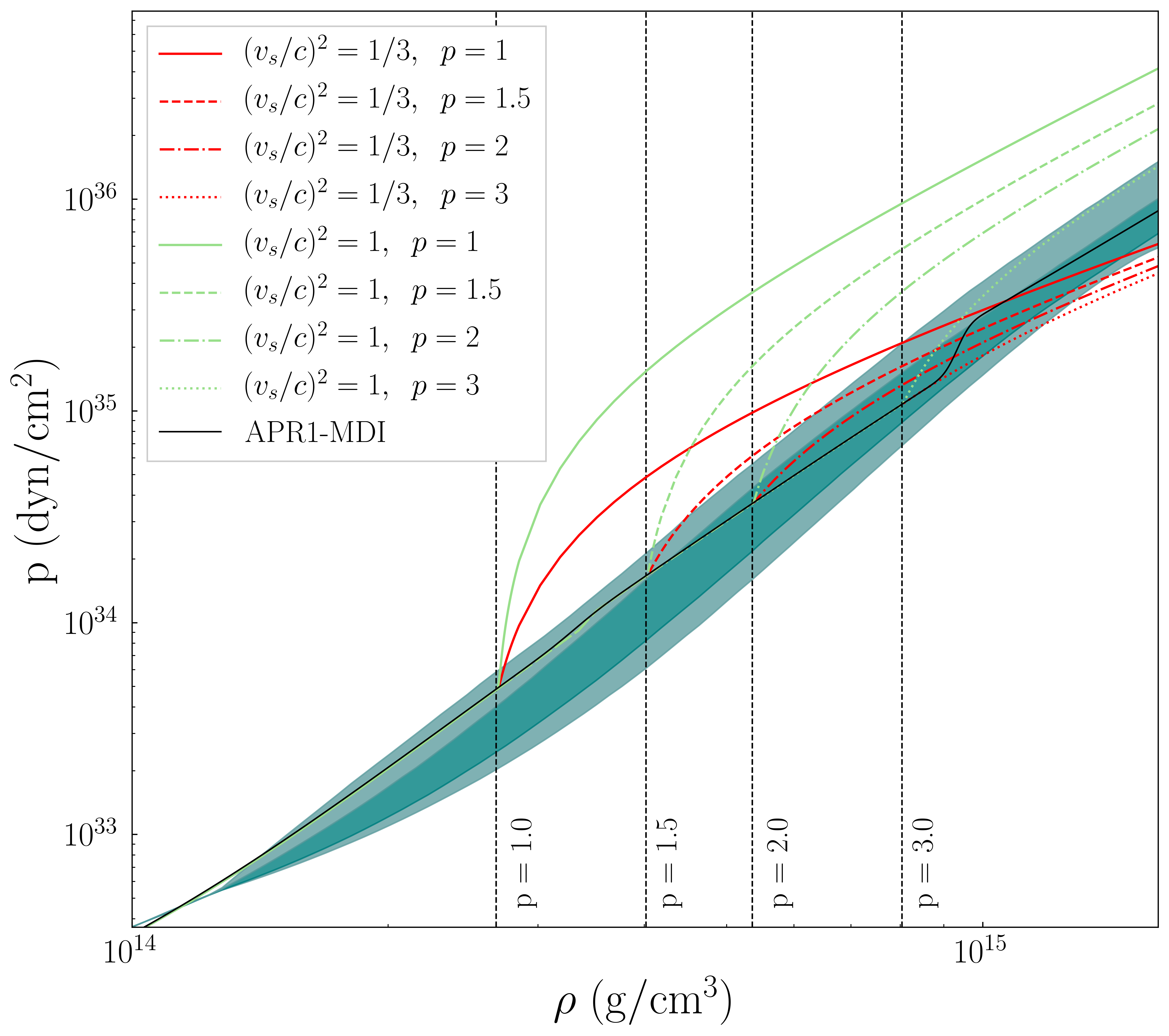}
\caption{Pressure vs rest mass density. The green (red) lines correspond to the upper (lower) limit for the speed of sound. Between the same kind of linestyle, the upper (lower) limit for the speed of sound coresponds to higher (lower) curves for each branch. The shaded region corresponds to the GW170817 estimation (data from Ref.~\cite{Abbott-2}). The solid black line corresponds to the APR1-MDI EoS~\cite{Akmal-98,Margaritis-2020}. The vertical black lines indicate the transition density $n_{{\rm tr}}$.}
\label{Prhosoundspeed}
\end{figure}
%%%%%%%%%%%%%%%%%%%%%%%%%%%%%%%%%%%%%%%%%%%%%%%%%%%%%%%%%%%%%%%%%%%%%%
%%%%%%%%%%%%%%%%%%%%%%%%%%%%%%%%%%%%%%%%%%%%%%%%%%%%%%%%%
%FIGURE-3
\begin{figure*}
\centering
\includegraphics[width=0.49\textwidth]{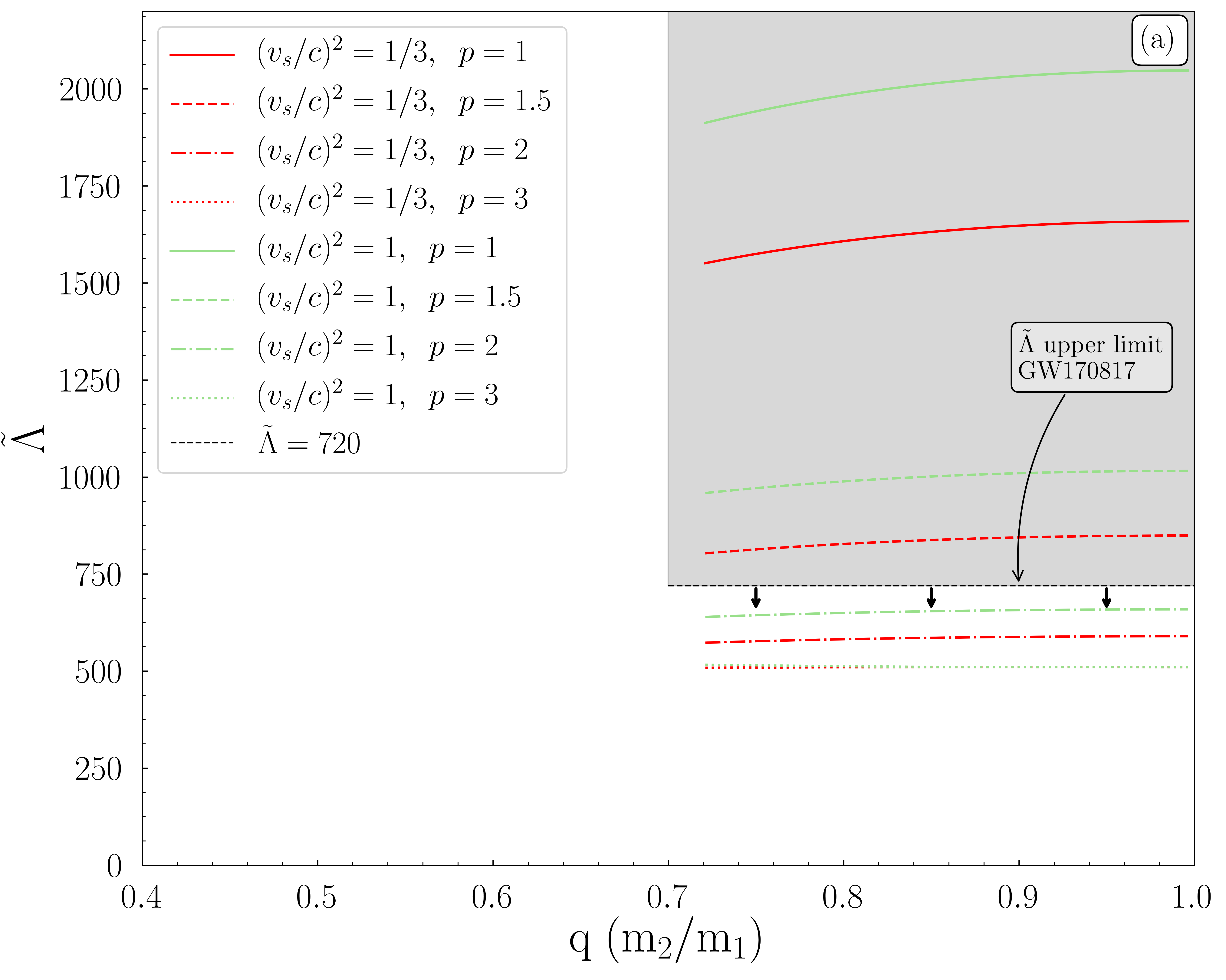}
~
\includegraphics[width=0.49\textwidth]{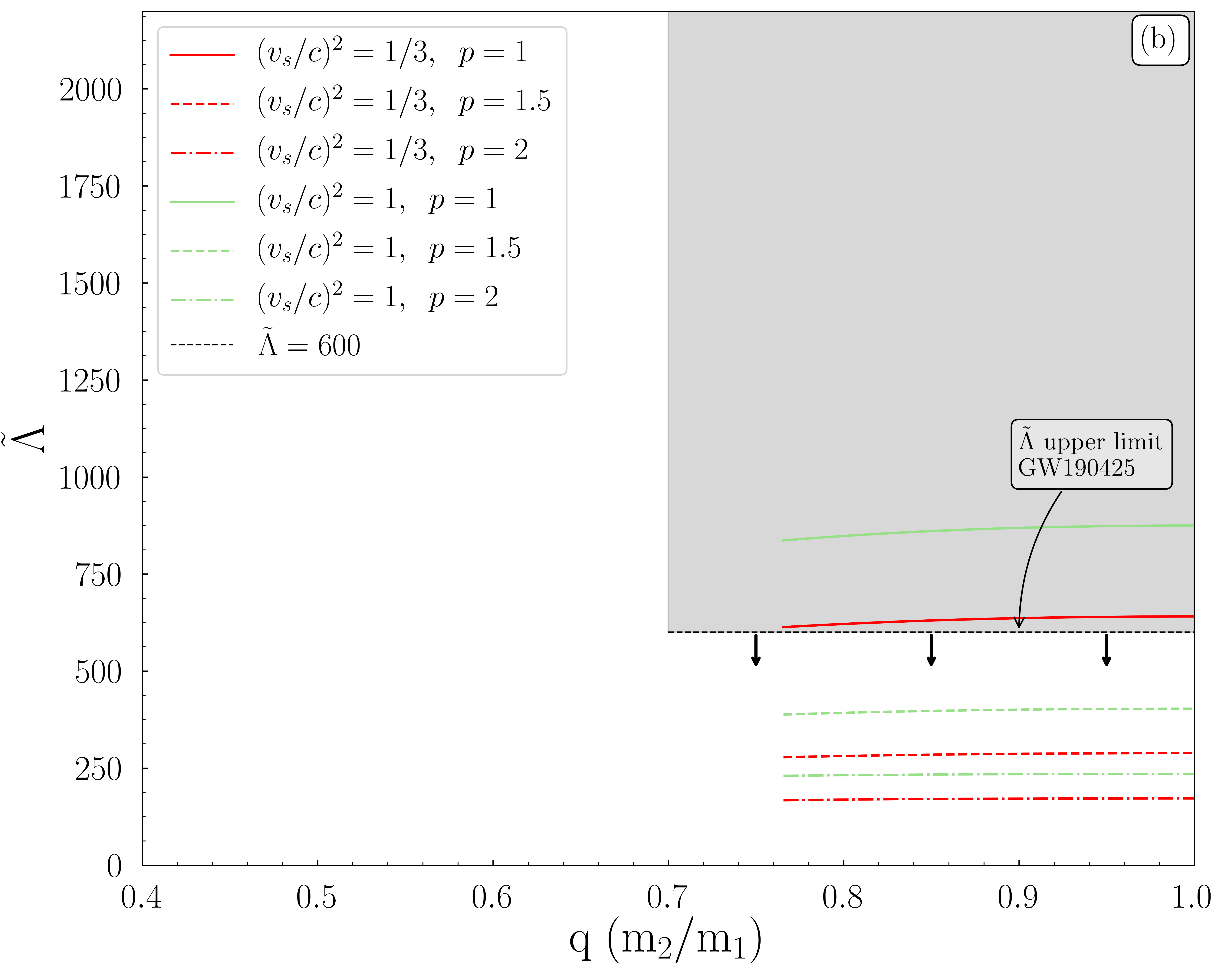}
\caption{The tidal deformability $\tilde{\Lambda}$ as a function of the binary mass ratio $q$ for the event (a) GW170817 and (b) GW190425. The corresponding upper observation limits for $\tilde{\Lambda}$ are also indicated, with the grey shaded region marking the excluded area. The red (green) lines correspond to the $(v_s/c)^2=1/3$ ($(v_s/c)^2=1$) limit.  Between the same kind of linestyle, the upper (lower) limit for the speed of sound coresponds to higher (lower) curves for each case. On the left panel, the black dashed line (with arrows) indicates the upper limit for $\tilde{\Lambda}$ from the reanalysis of the signal (see Ref.~\cite{Abbott-3}). On the right panel, the black dashed line (with arrows) corresponds to the upper bound for $\tilde{\Lambda}$~\cite{Abbott-4}.}
\label{Ltildeq}
\end{figure*}
%%%%%%%%%%%%%%%%%%%%%%%%%%%%%%%%%%%%%%%%%%%%%%%%%%%%%%%%%%%%%%%%%%%%%%
In Fig.~\ref{Prhosoundspeed} we display the constraints on pressure-rest mass density diagram form LIGO~\cite{Abbott-2} as well as the relevant predictions for the speed of sound bounds from the present work. To be more specific, the collaboration of Ref.~\cite{Abbott-2} used a spectral EOS parametrization in combination with the requirement that the EoS must support neutron stars up to at least $1.97\;M_{\odot}$. In Fig.~\ref{Prhosoundspeed}  the posterior analysis is indicated. According to Ref.~\cite{Abbott-2} the pressure posterior is shifted from the $90 \%$ credible prior region and towards the soft floor of the parametrized family of EoS. This means that the posterior is indicating support for softer EoS than the prior. In Fig.~\ref{Prhosoundspeed} is also displayed the implemented APR1-MDI EoS (which is the basis for the calculations) where the various branches correspond to the implementation of the bounds on the speed of sound. The change on the tread of the APR1-MDI EoS around $3.5 n_0$ is due to the parametrization of the MDI model in order to reproduce accurately the predictions of the APR1 EoS (for more details see Ref.~\cite{Koliogiannis-2019}). Obviously, the comparison leads to the main conclusion that any neutron star EoS that exhibits strong stiffness at low densities must be excluded. Moreover, the lower bound $c/\sqrt{3}$ must be reached only at densities higher than $1.5 n_0$ in order to be in accordance with the prediction of the LIGO. However, at higher  densities the lower  limit $c/\sqrt{3}$  must be violated in order to ensure stiff enough EoS. Actually, the present comparison supports the previous findings and conclusions. 

The case of binary neutron stars mergers offers a unique physical laboratory, suitable for studying the macroscopic properties of neutron stars (such as their mass, radius, tidal deformability and  moment of inertia) compared to their microscopic angle of view (EoS, pressure, density, nuclear parameters etc.). The effective tidal deformability, which is a binary combination of the tidal deformability of each component star, is a parameter which is appropriate to link the two scale approaches and contain information about the EoS (see Eq.~(\ref{L-tild-1})). This information is imprinted in the gravitational-wave signal as the leading-order of tidal effects in the waveform. Therefore, it can be measured by the gravitational-wave detectors~\cite{Abbott-1}. 

As it was mentioned above, our study takes into consideration the information from the gravitational wave emission of binary neutron stars coalescences. Hence, we used the constraints on the effective tidal deformability $\tilde{\Lambda}$, provided by the events GW170817 and GW190425 (we focus only on the low-spin scenario in order to be consistent with the known galactic binary neutron star systems, however, the chirp mass of the second event is not consistent with them~\cite{Abbott-4}). At first, the effective tidal deformability $\tilde{\Lambda}$ for the GW170817 event  was constrained as $\tilde{\Lambda}\leq800$ (corrected as $\tilde{\Lambda}\leq900$, with a chirp mass value of $\mathcal{M}_c=1.188\;M_\odot$~\cite{Abbott-1}), but after a reanalysis of the detection's data by LIGO, the constraint on $\tilde{\Lambda}$ estimated to be $\tilde{\Lambda}\leq720$, with a renewed value of chirp mass $\mathcal{M}_c=1.186\;M_\odot$, all at the $90\%$ conficence level~\cite{Abbott-3}. For the first event GW170817, we choose the component masses to vary within the range provided by LIGO, which means $m_1\in(1.36,1.60)\;M_\odot$ and $m_2\in(1.16,1.36)\;M_\odot$~\cite{Abbott-3}. To be more specific, by combining the range of one of the component masses and  Eq.~(\ref{chirpmass}), the other component mass can be determined. Hence, the binary mass ratio's range $q$ is known and therefore the effective tidal deformability, provided by Eq.~(\ref{L-tild-1}), can be presented as a function of the EoS, the binary mass ratio, and the individual tidal deformabilities $\tilde{\Lambda}=\tilde{\Lambda}(\Lambda_1,\Lambda_2,q;EoS)$.
%%%%%%%%%%%%%%%%%%%%%%%%%%%%%%%%%%%%%%%%%%%%%%%%%%%%%%%%%%%%%%%%%%%%%%
%FIGURE-4
\begin{figure*}
\centering
\includegraphics[width=0.49\textwidth]{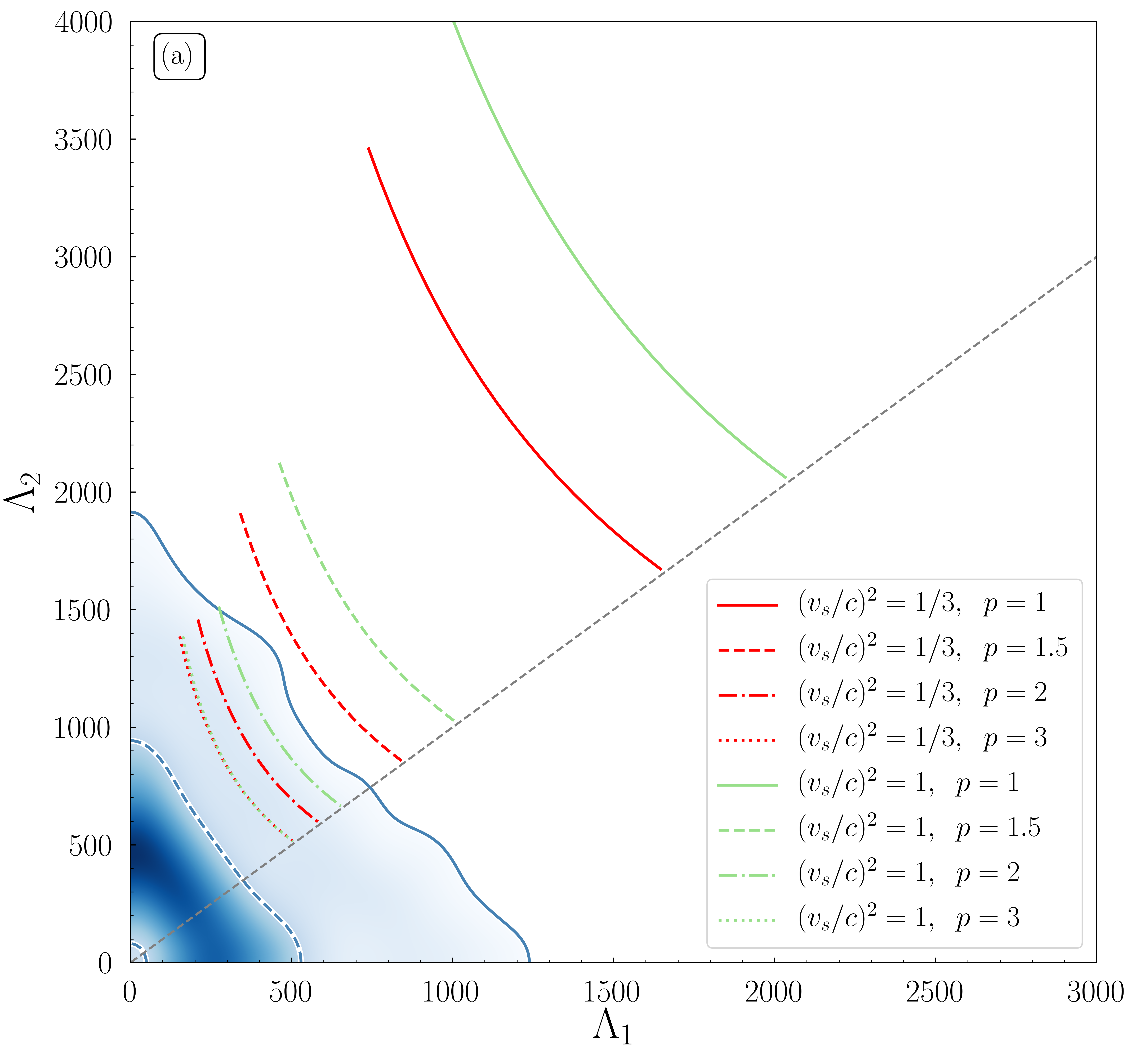}
~
\includegraphics[width=0.49\textwidth]{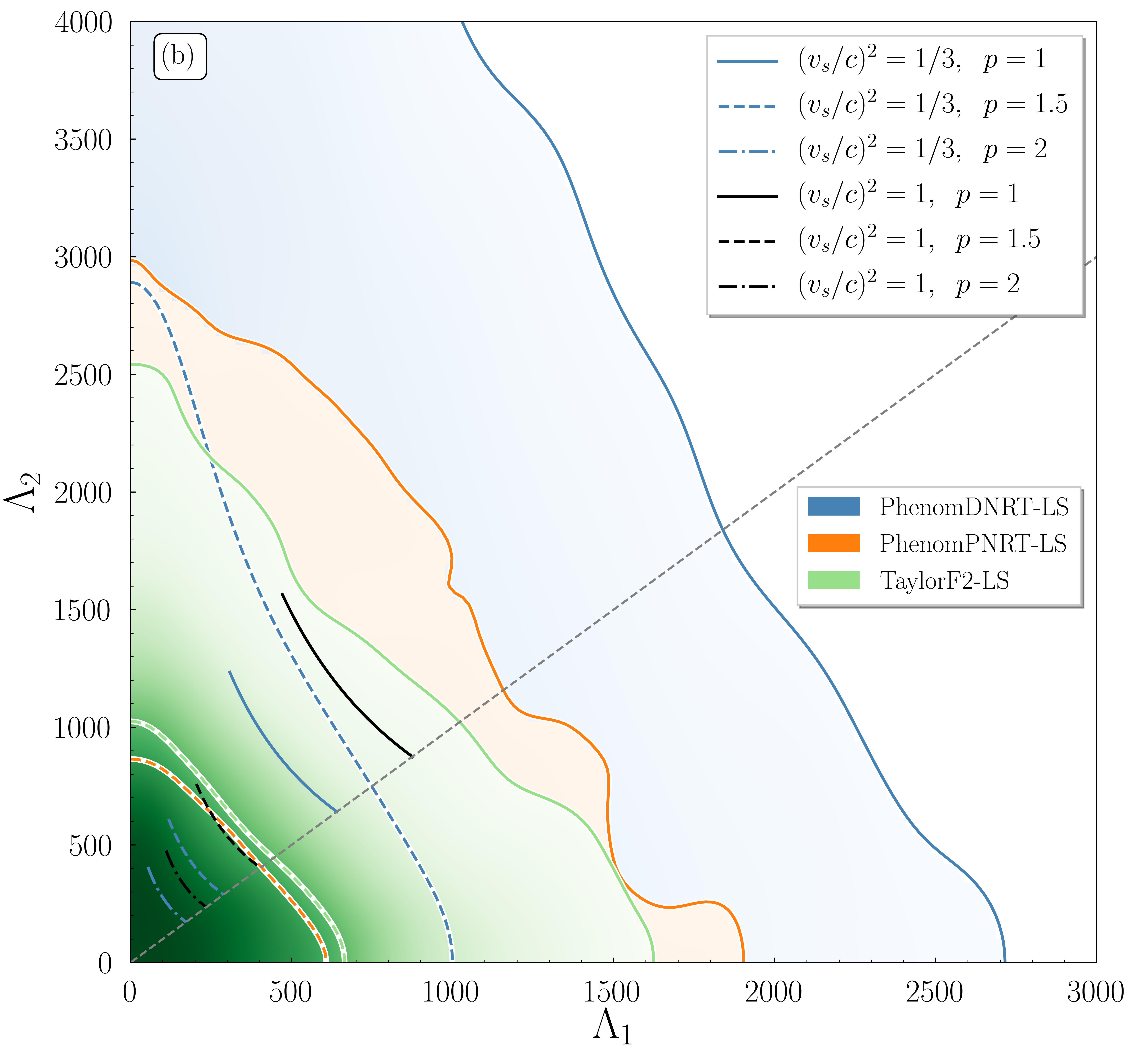}
\caption{$\Lambda_1-\Lambda_2$ diagram for the event (a) GW170817  and (b) GW190425. On the left panel, the shaded region corresponds to the posterior sample of the event~\cite{Abbott-3}. The solid (dashed) contour line indicates the $90\%$ ($50\%$) credible region. The grey dashed line corresponds to the $\Lambda_1=\Lambda_2=\tilde{\Lambda}$ case. The EoSs with $n_{{\rm tr}}=1,1.5\;n_0$ are beyond the credible regions. On the right panel, the shaded regions correspond to three different waveform models~\cite{Abbott-4}. The PhenomPNRT-LS model (orange) is the one we use for the upper limit on $\tilde{\Lambda}$. We notice that these posteriors disfavor low values of $\tilde{\Lambda}$, therefore their regions cannot provide appropriate constraints~\cite{Abbott-4}.}
\label{L1L2events}
\end{figure*}
%%%%%%%%%%%%%%%%%%%%%%%%%%%%%%%%%%%%%%%%%%%%%%%%%%%%%%%%%%%%%%%%%%%%%%

The effective tidal deformability $\tilde{\Lambda}$ is shown in Fig.~\ref{Ltildeq}, for both events. We notice that in the case of GW190425 event, the chirp mass was measured to be $\mathcal{M}_c=1.44\;M_\odot$~\cite{Abbott-4}, with the heavier component star's mass in our study to vary in the range $m_1\in(1.654,1.894)$ and the lighter's mass to be $m_2\in(1.45,1.654)$. Our choice for these values is based on the fact that even if the chirp mass is well determined by the gravitational-waves detectors, the range of the component masses has an overlap, and a strict common equal-mass bound has not been well defined~\cite{Abbott-4}. Therefore, we modify the mass range of the system in such a way that the binary mass ratio is $q\leq1$. Other estimations regarding to the range of masses were given in Ref.~\cite{Kilpatrick}. The upper limit for $\tilde{\Lambda}$, derived from GW190425 event, was determined as $\tilde{\Lambda}\leq600$~\cite{Abbott-4}. The diagram of $\tilde{\Lambda}$ as a function of $q$, in combination with the constraints on $\tilde{\Lambda}$ can highlight the possible constraints on the EoS. Especially, the upper limit on $\tilde{\Lambda}$ in Fig.~\ref{Ltildeq}(a), provided by GW170817,  leads to the exclusion of both cases of speed of sound for transition densities $n_{{\rm tr}}=1,1.5\;n_0$. By comparing this result to  Fig.~\ref{MR-1}, in which the transition density  $n_{{\rm tr}}=1.5n_0$ was not very clear if it is in agreement with the gravitational-waves observation, we  understood the extra tool that this diagram offered us. In particular, in Fig.~\ref{Ltildeq} the constraints on the upper limit of $\tilde{\Lambda}$ distinguish in a more efficient way the cases that must be excluded. Moreover, for the second event in Fig.~\ref{Ltildeq}(b), we observed that in general all  EoSs are shifted to lower values of $\tilde{\Lambda}$. This behavior is because of the higher value of chirp mass and the component masses. Contrary to the GW170817 event, the upper limit on $\tilde{\Lambda}$, provided by GW190425 event, excludes only the EoS with transistion density $n_{{\rm tr}}=n_0$, for both sound speed cases. For both events, the EoSs with higher values of transition density $n_{{\rm tr}}$ have smaller values of $\tilde{\Lambda}$. This means that the constraints on $\tilde{\Lambda}$ derived from gravitational-waves events favor softer EoSs. We mention here that for the GW190425 event, we removed from our study the cases with transition density $n_{{\rm tr}}=3\;n_0$ because of the fact that the EoS with $(v_s/c)^2=1/3$ and $n_{{\rm tr}}=3n_0$ can not reproduce the masses of this event. 

Another tool that is helpful for studying the EoS of dense matter through the tidal deformabilities, as derived from gravitational wave events, is the construction of $\Lambda_2-\Lambda_1$ space diagram. Similar to the previous study for the effective tidal deformability $\tilde{\Lambda}$ as function of $q$, we made the same assumptions for the chirp masses and the component masses for both events. The results for each EoS, combined with the posterior samples for each event, are displayed in Fig.~\ref{L1L2events}. 

In Fig.~\ref{L1L2events}(a) the blue shaded region indicates the posterior distribution for the event GW170817, taken from the reanalysis of GW170817 signal \cite{Abbott-3}. One can observe similar behavior with the $\tilde{\Lambda}-q$ diagram (Fig.~\ref{Ltildeq}(a)). We notice that the more stiff EoSs predict values beyond the credible region, contrary to the softer ones. In Fig.~\ref{L1L2events}(b), the shaded regions correspond to different posterior distributions for the GW190425~\cite{Abbott-4}. The difference on the posterior samples is due to the different waveform models that were applied in the signal analysis. As it was mentioned in Ref.~\cite{Abbott-4}, the effective tidal deformability $\tilde{\Lambda}$ is constrained to 1200 for the low-spin case, leading to the displayed upper limits for the contours. The explanation behind this behavior is the usage of a uniform prior distribution for $\tilde{\Lambda}$ which disfavors the lower values of $\tilde{\Lambda}$. By assuming a new flat prior for $\tilde{\Lambda}$, one can obtain the proper upper limit of $\tilde{\Lambda}\leq600$ that we used in the $\tilde{\Lambda}-q$ diagram. Although this more appropriate assumption is not visualized in Fig.~\ref{L1L2events}(b), the upper limit for $\tilde{\Lambda}$ that we used in our study concerning the event GW190425, is 600 and not 1200. Under the consideration of the $\tilde{\Lambda}\leq600$ bound, the constraints that the $\Lambda_1-\Lambda_2$ could impose in case of GW190425, are the same with the respective $\tilde{\Lambda}-q$ diagram.

In addition, we notice that the first event (GW170817) provides more stringent constraints on both the transition density $n_{{\rm tr}}$ and speed of sound $(v_s/c)^2$ in compare to the second event (GW190425). Therefore, we expect that binary neutron stars mergers with lower masses are more suitable for studying the EoS through the tidal deformability.

So far, we used the gravitational-wave detections of binary neutron stars mergers in order to broaden our knowledge regarding the speed of sound for the two borderline cases. The study of the effective tidal deformability $\tilde{\Lambda}$ as a function of the binary mass ratio $q$ has been a very helpful way to exploit the information on the upper limit of $\tilde{\Lambda}$, provided by the detection of gravitational-wave signal, in order to impose constraints on the EoS of dense matter and the speed of sound. At this point, the question that arises is the following: Is there a way to impose more stringent constraints on the speed of sound, by using the measured $\tilde{\Lambda}$, so that a specific limit on the possible values of speed of sound can be obtained? This was the idea that motivated our study. 

In order to answer the question, an alternative diagram to $\tilde{\Lambda}-q$ is needed. From Fig.~\ref{Ltildeq} one can see that the $\tilde{\Lambda}$ varies within a range of $(\tilde{\Lambda}_{{\rm min}},\tilde{\Lambda}_{{\rm max}})$ for each EoS. By combining this remark to the behavior of the M-R curves in Fig.~\ref{MR-1}, we observed that the wider variation on $\tilde{\Lambda}$ values for each EoS is connected with the bigger change in the radius (espacially the inclination of the $M(R)$ curve). Therefore, we studied the range on $\tilde{\Lambda}$  as a function of the transition density $n_{{\rm tr}}$ for each bound case of speed of sound, i.e. the $\tilde{\Lambda}_{{\rm min}}^{(1/3,1)}-n_{{\rm tr}}$ and $\tilde{\Lambda}_{{\rm max}}^{(1/3,1)}-n_{{\rm tr}}$ relations.

\begin{figure}
\centering
\includegraphics[width=0.49\textwidth]{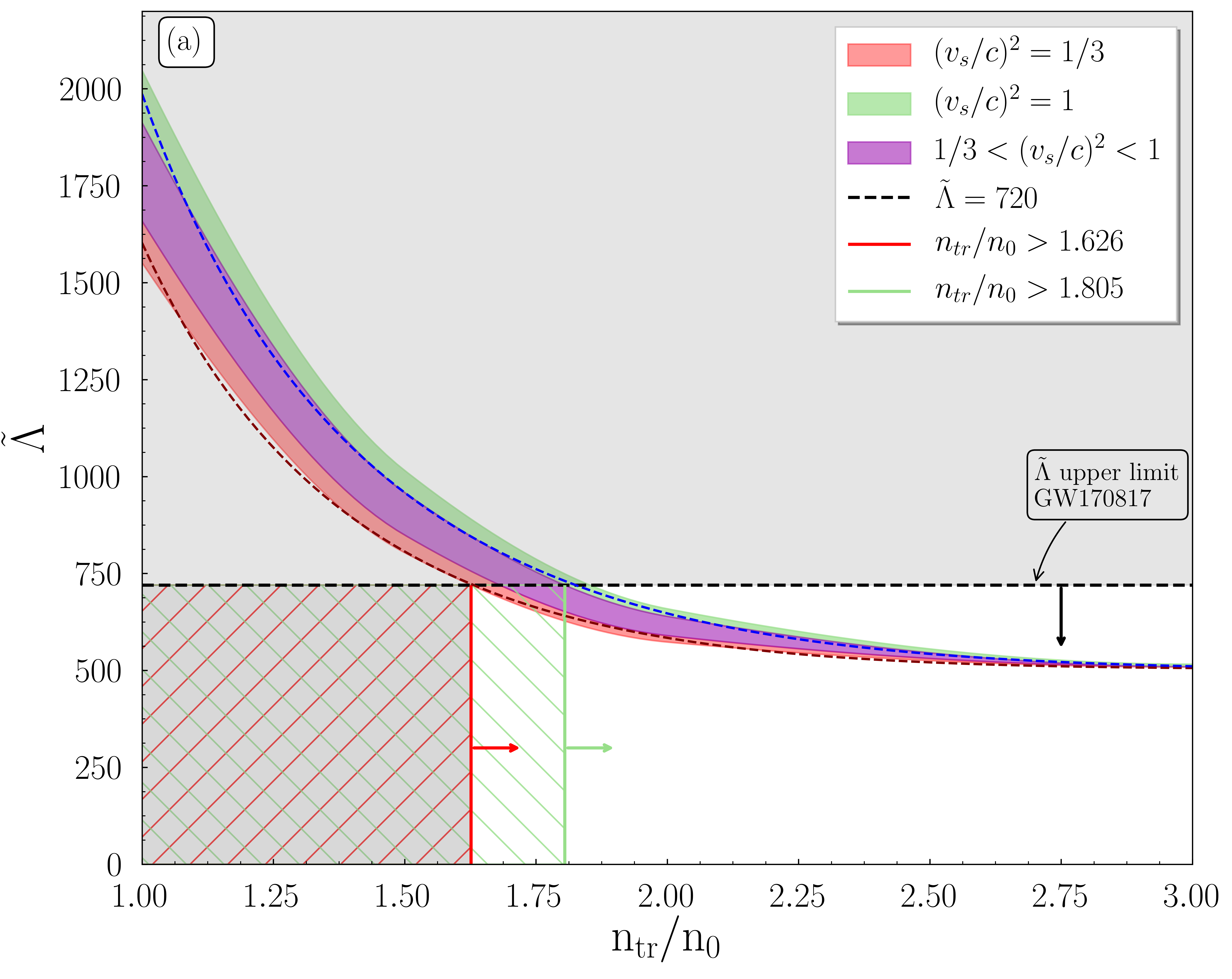}\\
\includegraphics[width=0.49\textwidth]{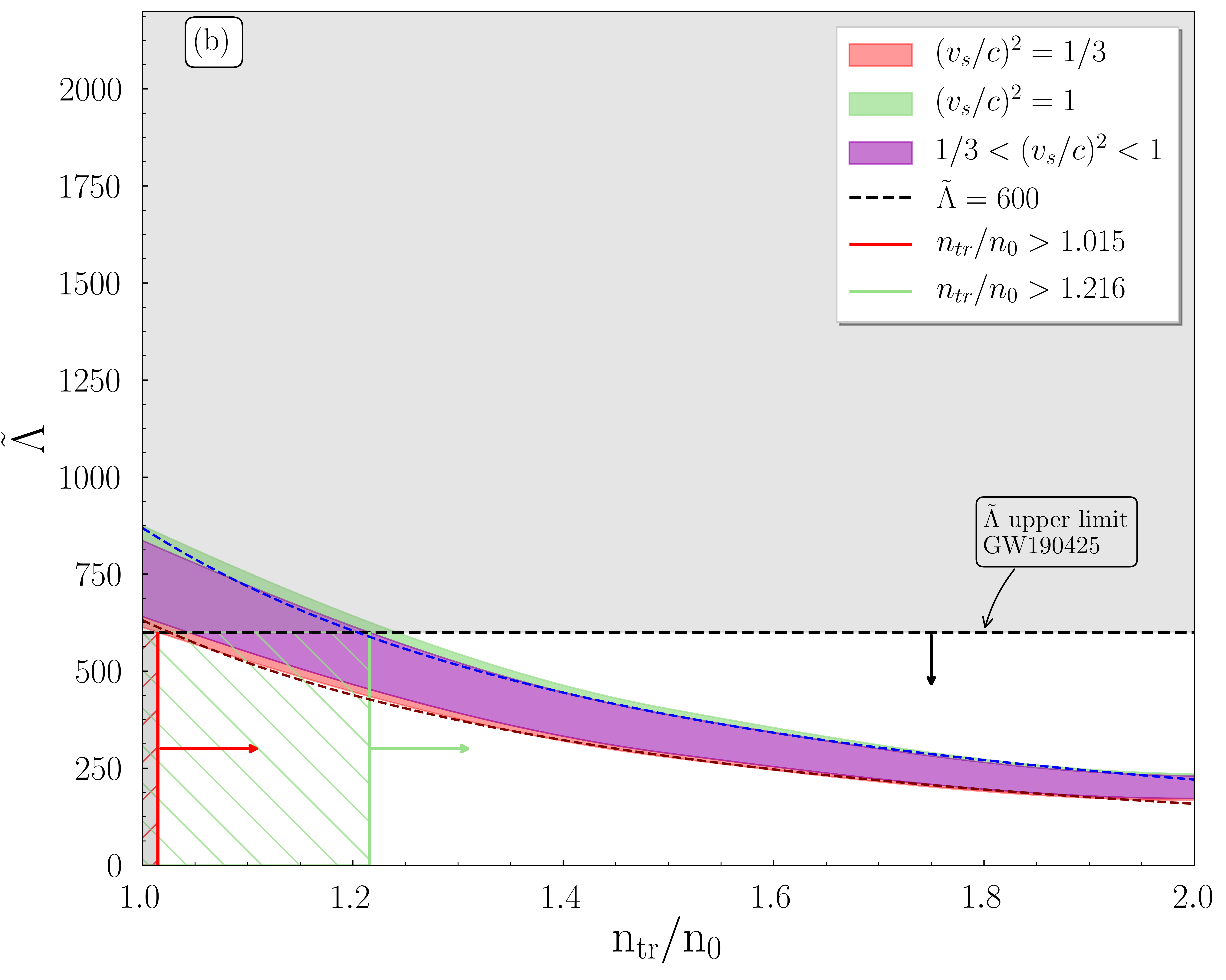}\\
\caption{Dependence of the effective tidal deformability $\tilde{\Lambda}$  on the transition density $n_{{\rm tr}}$ (in units of saturation density $n_0$) at the maximum mass configuration for the two  speed of sound bounds $v_s=c/\sqrt{3}$ and  $v_s=c$ and for the event (a) GW170817 (top panel) and (b) GW190425 (bottom panel). The corresponding upper observation limits for $\tilde{\Lambda}$~\cite{Abbott-3,Abbott-4}  as well as the compatible  lower transition density values are also indicated for both events.  The red (green) arrow marks the accepted region of transition density for the $v_s=c/\sqrt{3}$ ($v_s=c$) case. The red lower (green upper) curved shaded region corresponds to the $v_s=c/\sqrt{3}$ ($v_s=c$) limit. The purple intermediate shaded region indicates the predictions for middle values of speed of sound bound between the two limits of our study.}
\label{L-1}
\end{figure}

In Fig.~\ref{L-1} we display the dependence of the effective tidal deformability $\tilde{\Lambda}$  on the transition density $n_{{\rm tr}}$ at the maximum mass configuration for the two  speed
of sound bounds, $v_s=c/\sqrt{3}$ and  $v_s=c$, and the two events GW170817 (Fig.~\ref{L-1}(a))  and GW190425 (Fig.~\ref{L-1}(b)). The corresponding upper observational limits for $\tilde{\Lambda}$,  as well as the compatible  lower transition density values,  are also indicated in each case. Obviously the predictions on the bound which  considered  between the two bounds correspond to the middle region. 
%%%%%%%%%%%%%%%%%%%%%%%%%%%%%%%%%%%%%%%%%%%%%%%%%%%%%%%%%%%%%%%%%%%%%%
%FIGURE-5
%%%%%%%%%%%%%%%%%%%%%%%%%%%%%%%%%%%%%%%%%%%%%%%%%%%%%%%%%%%%%%%%%%%%%%
 
\begin{table*}
	\squeezetable
	\caption{Parameters of the Eq.~\eqref{fit1} and Eq.~\eqref{fit2} for both events and the various speed of sound bounds.}
	\begin{ruledtabular}
		\begin{tabular}{ccccccccc}
			\multirow{2}{*}{Speed of sound bounds} & \multicolumn{4}{c}{GW170817} & \multicolumn{4}{c}{GW190425} \\
			& $c_{1}$ & $c_{2}$  & $c_{3}$ & $c_{4}$ & $c_{1}$ & $c_{2}$ & $c_{3}$ & $c_{4}$   \\
			\hline
			$c$ & 500.835 & 0.258 & 53.457 & 0.873 & 47.821 & 0.055 &  10.651 & 1.068 \\			
			$c/\sqrt{3}$ & 503.115 & 0.325 & 38.991 & 1.493  & 43.195 & 0.069  & 5.024 & 1.950 \\
		\end{tabular}
	\end{ruledtabular}
	\label{tab:table1}
\end{table*}

One can observe that
\begin{enumerate}[(i)] 
\item The overall thickness decreases as the transition density $n_{{\rm tr}}$ grows. The explanation for such behavior lies in the variation of the radius $M(R)$ presented with the M-R diagram (see Fig.~\ref{MR-1}).
\item The thickness of each shaded region for the boundary cases that we study, decreases also as the $n_{{\rm tr}}$ gets larger values. The explanation is the same as the previous one (case (i)).
\item The shaded areas are shifted in the GW190425 event, compared with the GW170817 case, due to the increment of the masses (component and total) of the binary system. The same behavior was observed in Fig.~\ref{Ltildeq}(b), compared to  Fig.~\ref{Ltildeq}(a).
\end{enumerate}

In the case of the GW170817  event the lower limit for the transition density is $1.626 n_{\rm 0}$  for $v_s=c/\sqrt{3}$   and  $1.805 n_{\rm 0}$ for $v_s=c$. Correspondingly, for the second event  GW190425 the limits are $1.015 n_{\rm 0}$  for $v_s=c/\sqrt{3}$  and  $1.216 n_{\rm 0}$ for $v_s=c$. Obviously the first event imposes more stringent constraints on the EoS (see also below). In particular, the value of the speed of sound must be lower than $v_s=c/\sqrt{3}$, at least  up to  density $1.626 n_{\rm 0}$ (in order to keep the softness low enough to lead on the prediction of the tidal deformability). Moreover, the EoS must still remain casual at least up to density  $1.805 n_{\rm 0}$. Greif {\it et al.}~\cite{Greif-2019} remarked that according to the Fermi liquid theory (FLT) the speed of sound must be $v_{s,FLT}^2\leq0.163c^2$ for $n=1.5n_0$, meaning that the EoS cannot exceed this value for $n\leq1.5n_0$. This remark is in agreement with our finding of the lower limit $n_{{\rm tr}}=1.626n_0$ for the case of $v_s=c/\sqrt{3}$.

We notice that for the GW170817 event, the upper limit on $\tilde{\Lambda}$ can impose stringent constraints on the $n_{{\rm tr}}$, while a lower limit on $\tilde{\Lambda}$ could provide further information. Indeed, such a lower limit is provided both by the gravitational wave data~\cite{Abbott-2,Abbott-3} and the electromagnetic (EM) counterpart of the merger~\cite{Radice-1,Radice-2,Kiuchi-2019,Coughlin-2018,Coughlin-2019}. The usage of the EM counterpart as a source for a lower limit on $\tilde{\Lambda}>400$ proposed in Ref.~\cite{Radice-1}. Tews {\it et al.}~\cite{Tews-2018b} argued that this limit is not decisive  because EoSs with $\tilde{\Lambda}<400$ which predict even a neutron star with $M_{{\rm max}}=2.6\;M_\odot$, exist. Later, this value updated to be $\tilde{\Lambda}>300$~\cite{Radice-2}. Other studies suggested different lower values of $\tilde{\Lambda}_{{\rm min}}^{{\rm (EM)}}$~\cite{Coughlin-2018,Coughlin-2019}. Kiuchi {\it et al.}~\cite{Kiuchi-2019} articulated the dependence of the lower limit on $\tilde{\Lambda}$, derived by the EM counterpart of the binary neutron stars coalescences, from the binary mass ratio $q$, suggested a lower bound of $242$. The exact physical mechanism that describes the dependence of $\tilde{\Lambda}^{{\rm (EM)}}$ is still under investigation~\cite{Radice-1,Kiuchi-2019}. Most {\it et al.}~\cite{Most-2018} used this bound of Ref.~\cite{Radice-1} and demonstrated its significance in order to constrain further the tidal deformability $\tilde{\Lambda}_{1.4}$ and the radius $R_{1.4}$ of a $M=1.4\;M_\odot$ neutron star. For our case, a lower limit on $\tilde{\Lambda}$, similar to the proposed values, could not provide any further constraint, even if we consider the more optimistic boundary of $\tilde{\Lambda}\geq400$.

%(from the EM counterpart of the event) can provide additional information leading to the exclusion of the softer EoS~\cite{Radice-1,Shibata}. The exact lower limit on $\tilde{\Lambda}$ and the physical mechanism that describes the process is still under examination ~\cite{Radice-1,Shibata,Radice-2,Coughlin-2018,Coughlin-2019}. For our case, a lower limit on $\tilde{\Lambda}$, similar to the proposed values, could not provide any further constraint and thus could not be suitable. 

On the other hand, for the second event GW190425, the bigger masses that characterize it lead to smaller values on $\tilde{\Lambda}$, hence the upper limit on $\tilde{\Lambda}$ cannot provide any further constraints. On the contrary, we speculate that a lower limit on $\tilde{\Lambda}$ would be able to provide constraints, especially an upper limit for $n_{{\rm tr}}$. If this could be possible, then the binary neutron stars mergers with heavy component masses, would suitable to impose constraints to the upper limit of $n_{{\rm tr}}$ through the lower limit of $\tilde{\Lambda}$ as provided by the EM counterpart. Unfortunately, for the GW190425 event such a counterpart was not detected~\cite{Abbott-4,Kilpatrick}. 

In addition, we provide in Fig.~\ref{L-1} an expression for the $\tilde{\Lambda}_{{\rm min}}^{(1/3)}$ and $\tilde{\Lambda}_{{\rm min}}^{(1)}$ boundary curves of the red (lower) and green (upper) shaded regions, respectively. This expression provides the lower limit on $n_{{\rm tr}}^{(1/3)}$  and $n_{{\rm tr}}^{(1)}$, respectively.  The expression is given by the equation below, and the coefficients on the Table~\ref{tab:table1},
\begin{equation}
\tilde{\Lambda}=c_1\coth\Bigg[c_2\bigg(\frac{n_{{\rm tr}}}{n_0}\bigg)^{2}\Bigg].
\label{fit1}
\end{equation}

Tews {\it et al.} ~\cite{Tews-2018,Tews-2019} noticed (using the Chiral effective field theory) that the EoS of neutron matter up to twice the saturation density, requires the violation of the conformal limit $(v_s/c)^2<1/3$ so that would be stiff enough to provide a $2\;M_\odot$ neutron star. Therefore, the behavior of the speed of sound for intermediate densities should be non-monotonous. Regarding the connection between $v_s^2$ and $M_{{\rm max}}$, the highest mass is provided by the stiffest EoSs, i.e. the higher value of speed of sound. For a neutron star with $M_{{\rm max}}=2\;M_\odot$ at $n_{{\rm tr}}=2n_0$, the lower bound for the maximum value of speed of sound estimated to be $v_s^2\geq0.4c^2$~\cite{Tews-2018,Tews-2019}. In a  very recent study of Tews {\it et al.}~\cite{Tews-2020} for the GW190814 event, it was mentioned that the maximum speed of sound must be $v_s^2\geq0.6c^2$ in order to provide a stiff enough EoS.

\begin{figure}
\centering
\includegraphics[width=0.49\textwidth]{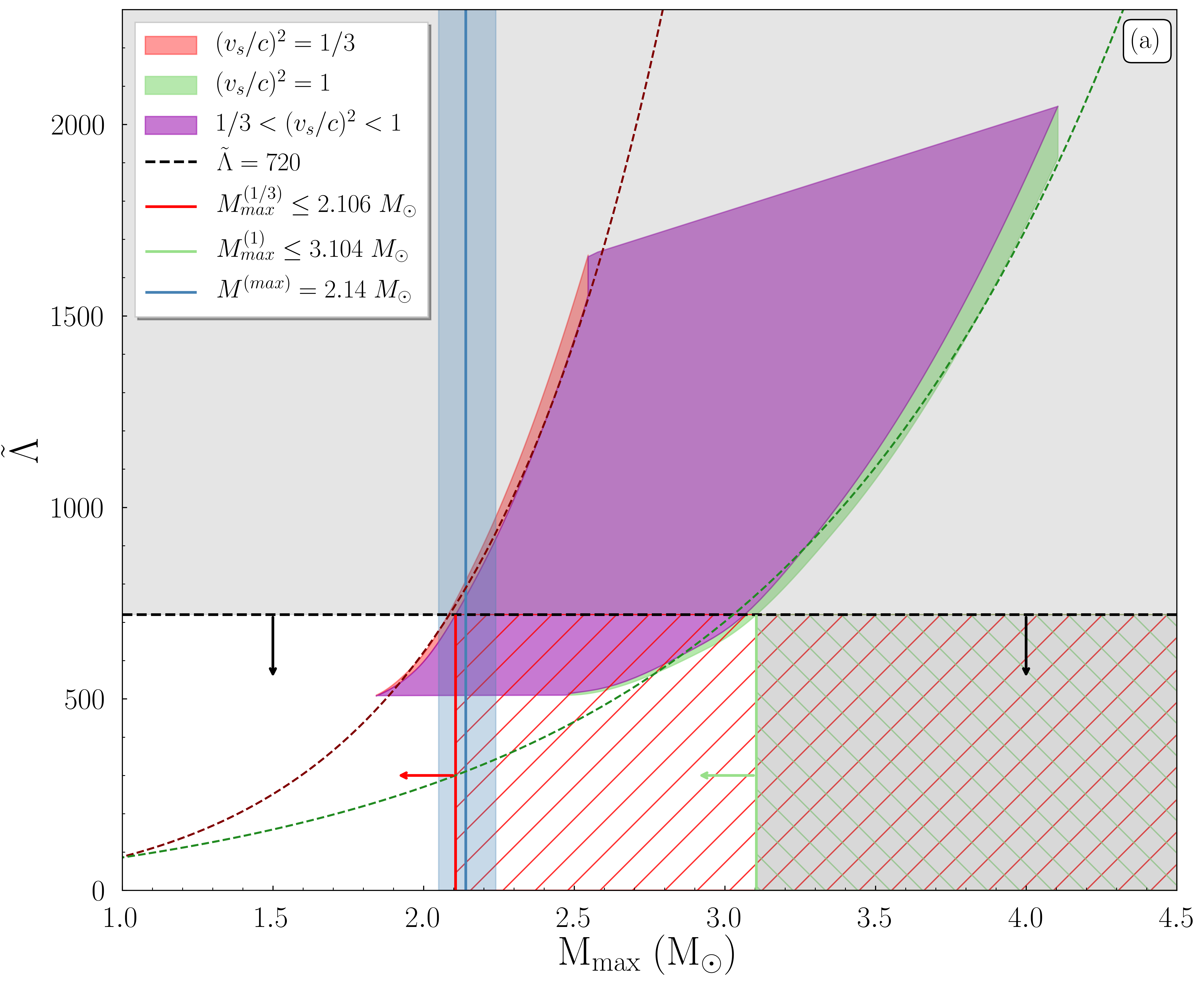}\\
\includegraphics[width=0.49\textwidth]{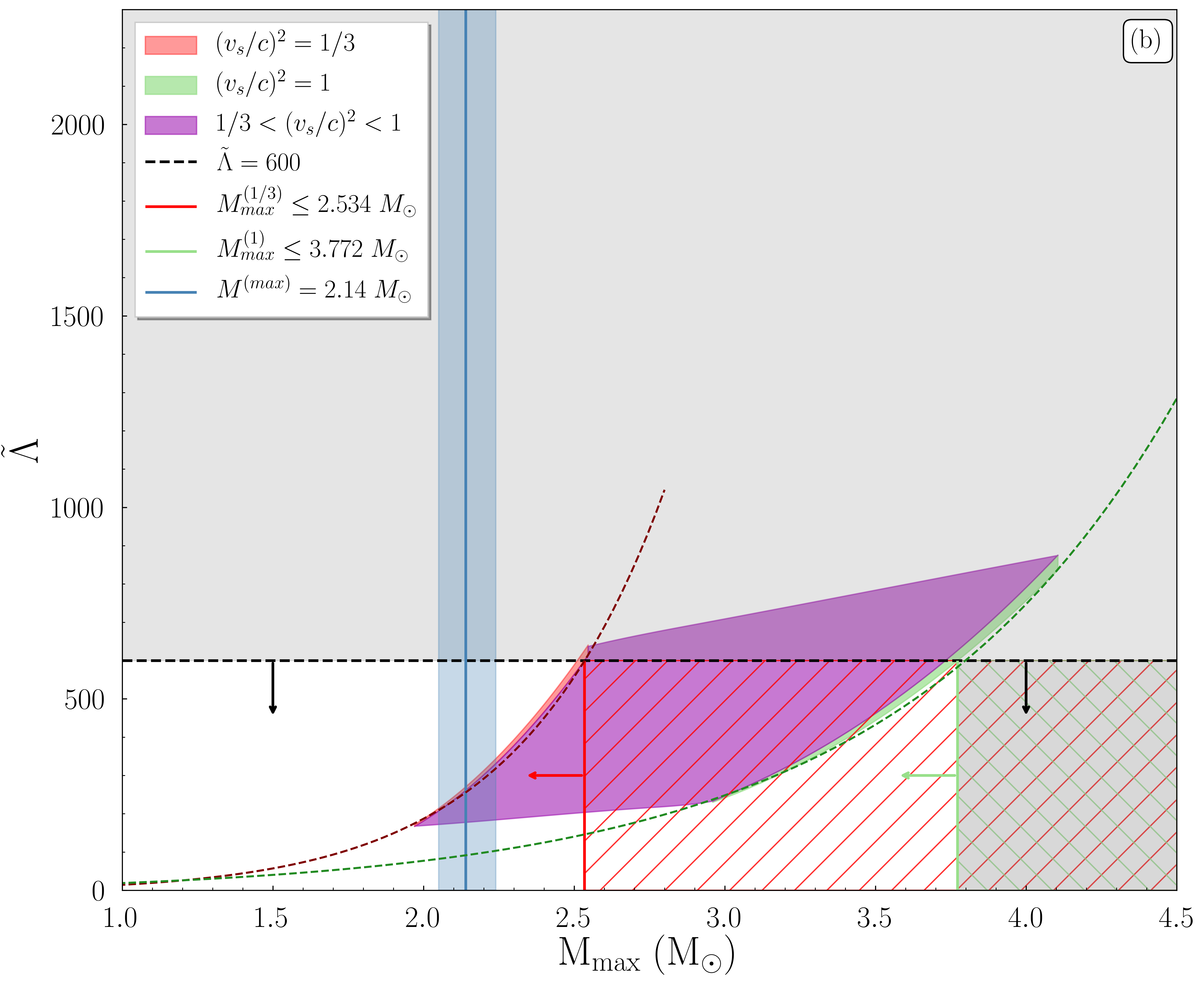}\\
\caption{The effective tidal deformability $\tilde{\Lambda}$ as a function of the maximum mass for the two  speed of sound bounds $v_s=c/\sqrt{3}$ and  $v_s=c$ and for the event (a) GW170817   (top panel) and (b) GW190425 (bottom panel). The corresponding upper observation limits for $\tilde{\Lambda}$ (black dashed lines with arrows, see Refs.~\cite{Abbott-3,Abbott-4}), the compatible maximum mass shaded regions, for $v_s=c/\sqrt{3}$ (left red) case, $v_s=c$ case (right green) and the middle cases (purple), as well  the current observed maximum neutron star mass  $M=2.14_{-0.09}^{+0.10}\;M_{\odot}$ (blue shaded vertical region, see Ref.~\cite{Cromartie-2019}) are also indicated. The red left (green right) arrow marks the accepted region of maximum mass $M_{{\rm max}}$ for $v_s=c/\sqrt{3}$ ($v_s=c$) case. The green (red) curved dashed line describes the fitted Eq.~(\ref{fit2}).}
\label{L-2}
\end{figure}

Subsequently, similar to the previous process which led to Fig.~\ref{L-1}, we constructed in Fig.~\ref{L-2} the range of $\tilde{\Lambda}$ (shaded regions) as a function of the maximum mass that each EoS provides. The goal is to use the effective tidal deformability $\tilde{\Lambda}$ as a variety of $(\tilde{\Lambda}_{{\rm min}},\tilde{\Lambda}_{{\rm max}})$ in order to impose constraints on the maximum mass for each boundary case of sound speed.

In Fig.~\ref{L-2} we display the dependence of the effective tidal deformability $\tilde{\Lambda}$ as a function of the maximum mass for the two speed of sound bounds and for both events. The corresponding upper observational limit for $\tilde{\Lambda}$ (black dashed horizontal line), the compatible maximum mass in each case (horizontal arrows), as well as the current observed maximum neutron star mass  $M=2.14_{-0.09}^{+0.10}\;M_{\odot}$ (shaded blue vertical region) are also indicated .
%%%%%%%%%%%%%%%%%%%%%%%%%%%%%%%%%%%%%%%%%%%%%%%%%%%%%%%%%%%%%%%%%%%%%%
%FIGURE-6
%%%%%%%%%%%%%%%%%%%%%%%%%%%%%%%%%%%%%%%%%%%%%%%%%%%%%%%%%%%%%%%%%%%%%%

%\begin{table*}
%	\squeezetable
%	\caption{Parameters of the Eq.~\eqref{fit2} for both events and the various speed of sound bounds.}
%	\begin{ruledtabular}
%		\begin{tabular}{ccccc}
%			\multirow{2}{*}{Speed of sound bounds} & \multicolumn{2}{c}{GW170817} & \multicolumn{2}{c}{GW190425} \\
%			& $c_{1}$ & $c_{2}$ & $c_{1}$ & $c_{2}$   \\
%			\hline
%			$c$ & 53.457 & 0.873 &  10.651 & 1.068  \\
%			
%			$c/\sqrt{3}$ & 38.991 & 1.493 & 5.024 & 1.950 \\
%		\end{tabular}
%	\end{ruledtabular}
%	\label{tab:table2}
%\end{table*} 
In Fig.~\ref{L-2} is clearly displayed the strong tension between the predicted maximum mass and the upper limit of the observed $\tilde{\Lambda}$. For the first event the upper bound of  $\tilde{\Lambda}$ is compatible with a maximum mass value $2.106\;M_{\odot}$ for $v_s=c/\sqrt{3}$ and  $3.104\;M_{\odot}$ for $v_s=c$.  However, this limit corresponds to transition density close to the value $1.5 n_{\rm 0}$. Experimental evidence or estimations are against this value. Consequently, the simultaneously derivation of the maximum mass combined with the experimental knowledge that the EoS cannot take this bound of sound speed for $n_{{\rm tr}}=1.5n_0$, are in contradiction. The upper limit on $M_{{\rm max}}$ for the case of $(v_s/c)^2=1/3$ lays roughly inside the estimation of the observed maximum mass. There are two different points of view that antagonize one another. The constraints derived by the upper limit on $\tilde{\Lambda}$ lead to more soft EoSs, contrary to the observational estimations of the maximum mass of neutron stars which lead to more stiff EoSs. This difference becomes less contradictory as the speed of sound takes larger values, with the causal scenario of $v_s=c$ leading to a very wide area for the maximum mass.

In the case of the second event GW190425, the constraints provided by the measured $\tilde{\Lambda}$ are less stringent, with a maximum mass value of $M_{{\rm max}}\leq2.534\;M_\odot$ for $v_s=c/\sqrt{3}$ and $M_{{\rm max}}\leq3.772\;M_\odot$ for $v_s=c$. However, as we mentioned in the $\tilde{\Lambda}-n_{{\rm tr}}$ diagram, a possible lower limit on $\tilde{\Lambda}$ for events with big component masses such as GW190425, could lead to constraints on the lower maximum mass. 

In addition, we used the data in order to provide an expression that describes the $\tilde{\Lambda}$ as a function of the maximum mass $M_{{\rm max}}$. The expression is given by the following equation and the coefficients on the Table~\ref{tab:table1},
\begin{equation}
\tilde{\Lambda}=c_3\bigg(e^{M_{{\rm max}}}-1\bigg)^{c_4}.
\label{fit2}
\end{equation}

Various studies suggest an upper limit on the possible maximum mass $M_{{\rm max}}$ of a neutron star, based on the GW170817 event~\cite{Rezzolla-2018,Magalit-2017,Ruiz-2018,Shibata-2017,Shibata-2019,Shao-2020}. Raaijmakers {\it et al.}~\cite{Raaijmakers-2020} by using a joint analysis of NICER and GW170817 detections, estimated the maximum mass of a neutron star, for two parameterizations: (a) $M_{{\rm max}}=2.26^{+0.16}_{-0.24}\;M_\odot$ (polytropic model) and (b) $M_{{\rm max}}=2.13^{+0.26}_{-0.22}\;M_\odot$ (speed of sound model). By adapting such an upper bound on the $M_{{\rm max}}$ in Fig.~\ref{L-2}, there is an additional constraint on the behavior of the speed of sound. To be more specific, by taking into consideration the estimated upper limit $M_{{\rm max}}\leq2.33\;M_\odot$~\cite{Rezzolla-2018}, the case of $(v_s/c)^2=1$ in Fig.~\ref{L-2}(a) for the GW170817 event should be excluded. On the contrary, the estimated upper limit $M_{{\rm max}}\leq2.106\;M_\odot$ for the $(v_s/c)^2=1/3$ bound,  is a more tight constraint. Moreover, an upper limit such as $M_{{\rm max}}\leq2.33\;M_\odot$ places a general upper bound on the possible intermediate values of speed of sound (intermediate shaded area in the figure). As for the second event in Fig.\ref{L-2}(b), a decisive upper limit on $M_{{\rm max}}$ could constrain even the $(v_s/c)^2=1/3$ case.

\begin{figure}
\centering
\includegraphics[width=0.49\textwidth]{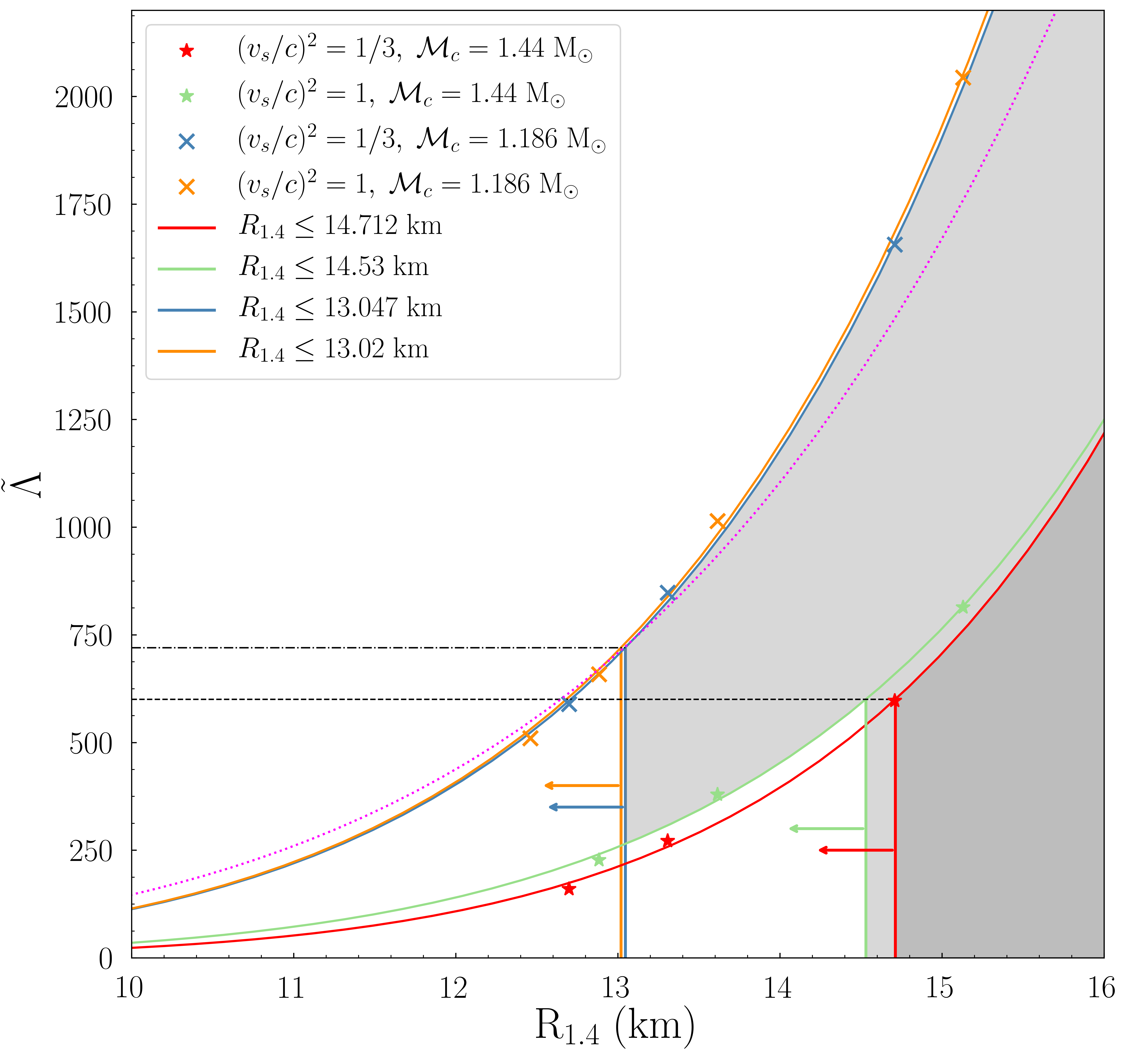}
\caption{The tidal deformability $\tilde{\Lambda}$ as a function of $R_{1.4}$ for both events and boundary cases of speed of sound. The dashed (dashdotted) horizontal black line corresponds to the upper limit on $\tilde{\Lambda}$ for GW190425 (GW170817) event, taken from Refs.~\cite{Abbott-3,Abbott-4}. The grey shaded regions show the excluded areas. The arrows indicate the allowed values of $R_{1.4}$ for each case. The purple dotted curve demonstrates the proposed expression by Refs.~\cite{Zhao,SoumiDe}.}
\label{LR}
\end{figure}

Lastly, we study the relation between the $\tilde{\Lambda}$ and the radius of a $1.4\;M_\odot$ neutron star, for both events. In Fig.~\ref{LR} we display the  dependence of  $\tilde{\Lambda}$  on the radius $R_{1.4}$ of a neutron star with mass $M=1.4\;M_\odot$. At first sight, the combination of the upper limit on $\tilde{\Lambda}$ with the speed of sound bound leads to a limitation on the maximum values of the radius, especially in the case of $v_s=c/\sqrt{3}$. Moreover, there is a trend between $\tilde{\Lambda}$ and $R_{1.4}$, which was remarked also by Raithel {\it et al.}~\cite{Raithel-2018}, mentioning that the effective tidal deformability depends strongly on the radii of the stars rather on the component masses. This strong dependence is presence in Fig.~\ref{LR}. 
%%%%%%%%%%%%%%%%%%%%%%%%%%%%%%%%%%%%%%%%%%%%%%%%%%%%%%%%%%%%%%%%%%%%%%
%FIGURE-7
%%%%%%%%%%%%%%%%%%%%%%%%%%%%%%%%%%%%%%%%%%%%%%%%%%%%%%%%%%%%%%%%%%%%%%

In particular, for the GW170817 event the fit curves of the two limited cases, blue (orange) for the $(v_s/c)^2=1/3$ ($(v_s/c)^2=1$) bound of the speed of sound, are almost identical. The cross marks indicate the specific values for each case. Since we considered 4 cases of transition density $n_{{\rm tr}}$, the total number of marks is expected to be 8. We notice that for $n_{{\rm tr}}=3n_0$, the two limits of $v_s$ predict identical values (orange cross). This is well understood by comparing it with their behavior in Fig.~\ref{MR-1} in which, for the mass range of GW170817 event, their M-R curves are identical. In addition, one can observe that for bigger values of $\tilde{\Lambda}$ the distance between the predicted values of each case is getting bigger. This behavior is in agreement with the general behavior of $\tilde{\Lambda}-q$ curves of Fig.~\ref{Ltildeq}(a), in which the distance between the curves increases for bigger values of $\tilde{\Lambda}$. Since the increment is related to the $n_{{\rm tr}}$, the differentiation for small values of $n_{{\rm tr}}$  is more obvious, i.e. the effect of each sound speed's bound case is easier to be manifested. The dotted purple line indicates the approximate relation of Refs.~\cite{Zhao,SoumiDe}. We underline that this relation is valid only for the first event and for specific assumptions on the components' radii. In particular, the main assumption of the provided expression consists on the $R_1\approx R_2$  relation. By comparing the curves' behavior in Fig.~\ref{LR} to the M-R curves of Fig~\ref{MR-1}, we can see that for smaller values of $n_{{\rm tr}}$, i.e. stiffer EoS, (a) the inclination of the curves increases and (b) the differentiation between the M-R curves of boundary cases is getting bigger. By taking into consideration the strong dependence of $\Lambda_i$ from $R$ (see Eq.~(\ref{Lamb-1})), combined with the previous remarks, the deviation of the fitting expression for bigger values of $\tilde{\Lambda}$ can be interpeted. 

The grey shaded area corresponds to the ecxluded region due to the upper limit on $\tilde{\Lambda}$, provided by Ref.~\cite{Abbott-3}. This upper limit on $\tilde{\Lambda}$ leads to constraints on the radius $R_{1.4}$, especially $R_{1.4}\leq13.047{\rm\;km}$ for $(v_s/c)^2=1/3$ bound and $R_{1.4}\leq13.02{\rm\;km}$ for $(v_s/c)^2=1$ bound. These upper limits are consisent to other analyses~\cite{Abbott-2,SoumiDe,Coughlin-2019,Tews-2018b,Most-2018,Raithel-2018,Annala-2018,Fasano-2019,Fattoyev-2018}. By using the gravitational wave and electromagnetic parts of the GW170817 event, Burgio {\it et al.}~\cite{Burgio-2018} constrained the radious of a $1.5\;M_\odot$ neutron star in the range $11.8{\rm\;km}\lesssim R_{1.5}\lesssim13.1{\rm\;km}$. Combining multimessenger information of the GW170817 merger with the Chiral effective theory, Capano {\it et al.}~\cite{Capano-2019} found that for a density up to twice the saturation density $2n_0$ the radius of a $1.4\;M_\odot$ neutron star is $R_{1.4}=11.0^{+0.9}_{-0.6}{\rm\;km}$. A multimessenger based constrain, using the recent observation of the isolated pulsar PSR $J0030+0451$, led to $R_{1.4}=12.1^{+1.2}_{-0.8}{\rm\;km}$~\cite{Jiang-2020}. Recently, a nonparametric-based approach yielded $R_{1.4}=12.51^{+1.00}_{-0.88}\;{\rm km}$~\cite{Essick-2020a}. Additionally, a recent work combining the Chiral effective field theory at low densities with observational data from gravitational waves (GW170817), pulsars and NICER, estimated the radius of a $1.4\;M_\odot$ neutron star to be $R_{1.4}=12.54^{+0.71}_{-0.63}{\rm\;km}$ with a maximum mass $M_{{\rm max}}=2.24^{+0.31}_{-0.23}\;M_\odot$~\cite{Essick-2020b}.

Moving on to the second event, we choosed the $R_{1.4}$ as a reference for consistency with the GW170817 event. We notice that the exact mass range of GW190425 event is still under examination ~\cite{Kilpatrick}. The red (green) lines and marks correspond to the $(v_s/c)^2=1/3$ ($(v_s/c)^2=1$) bound case. The shaded grey area indicates the excluded region as a result of the upper limit on $\tilde{\Lambda}$ \cite{Abbott-4}. The red and green arrow indicates the allowed region for each case. For $(v_s/c)^2=1/3$ the constraint on the radius is $R_{1.4}\leq14.712{\rm\;km}$, while for $(v_s/c)^2=1$ is $R_{1.4}\leq14.53{\rm\;km}$. These constraints are more stringent comparing to the $15{\rm\;km}$ and $16{\rm\;km}$ of Ref.~\cite{Abbott-4}.  Dietrich {\it et al.}~\cite{Dietrich-2020} performed a multi-messenger analysis of GW170817 event, in combination to the recent and less informative GW190425, leading to a more tight constraint on the $R_{1.4}$ radius $R_{1.4}=11.74^{+0.98}_{-0.79}{\rm\;km}$. Lastly, Landry {\it et al.}~\cite{Landry-2020} using an nonparametric approach, considering both events (see also Ref.~\cite{Essick-2020a}), concluded to the estimated value $R_{1.4}=12.32^{+1.09}_{-1.47}{\rm\;km}$. Furthermore, it was found that the joint contribution of gravitational wave and NICER data favor the violation of the conformal limit $(v_s/c)^2<1/3$. It was remarked that the constraints on the speed of sound, derived by the joint analysis, suggest  the violation of the conformal limit around  $~4\rho_{{\rm nuc}}$ density, where $\rho_{{\rm nuc}}=2.8\times10^{14}\;{\rm g/cm^3}$ is the nuclear saturation density~\cite{Landry-2020}.

Moreover, one can observe that the curves behave similar to the first event. For bigger values of $n_{{\rm tr}}$ the distance between the points is getting bigger. One of the main difference between the two events, is that for the second one, the curves and the points have been shifted to smaller values of $\tilde{\Lambda}$. This is due to the bigger chirp mass $\mathcal{M}_c$. Also, the fitting lines are more distincted from each other, contrary to the GW170817 event in which they were almost identical; but there is a common trend (see also Ref.~\cite{Raithel-2018}). We applied the following fitting expression
\begin{equation}
\tilde{\Lambda}=c_1R_{1.4}^{c_2},
\label{fitR14}
\end{equation}
where $R_{1.4}$ is in ${\rm km}$, similar to the proposed relations of Refs.~\cite{Zhao,SoumiDe,Tews-2019}. The coefficients for each case are given in Table~\ref{tab:table3}.
\begin{table}[H]
	\squeezetable
	\caption{Coefficients of Eq.~\eqref{fitR14} for the two speed of sound bounds.}
	\begin{ruledtabular}
		\begin{tabular}{cccc}
			Event & Speed of sound bounds & $c_{1}$ & $c_{2}$ \\
			\hline\multirow{2}{*}{GW170817} & $c$ & $0.12357\times10^{-4}$ & 6.967 \\
			& $c/\sqrt{3}$ & $0.12179\times10^{-4}$ & 6.967 \\
			\multirow{2}{*}{GW190425} & $c$ & $0.870\times10^{-6}$ & 7.605 \\
			& $c/\sqrt{3}$ & $0.088\times10^{-6}$ & 8.422 \\
		\end{tabular}
	\end{ruledtabular}
	\label{tab:table3}
\end{table}
We notice that rigorous measurements, or at least constraints, on $R_{1.4}$ may greatly help  to gain useful  insights   on the speed of sound bounds. 

%%%%%%%%%%%%%%%%%%%%%%%%%%%%%%%%%%%%%%%%%%%%%%%%
\section{Concluding remarks}
%%%%%%%%%%%%%%%%%%%%%%%%%%%%%%%%%%%%%%%%%%%%%%%%%
In this work we studied possible constraints on the speed of sound (stiffness of the EoS) and transition density $n_{{\rm tr}}$ which are based on the very recent observations of the events GW170817   and GW190425. The method which was implemented involved mainly the upper limits of the effective tidal deformability $\tilde{\Lambda}$ (estimated from the mentioned events), combined with measurements and estimations of the maximum neutron star mass. As basis of our study, we used the APR1-MDI EoS, for two boundary cases of speed of sound~\cite{Koliogiannis-2019,Margaritis-2020}; the lower bound of $v_s=c/\sqrt{3}$ and the upper one of $v_s=c$. 

Firstly, we examined the behavior of EoSs for each case, by using the information on the component masses of GW170817 event in combination with the NICER's data. The M-R diagram was able to impose some robust constraints, by excluding the more stiff EoSs with $n_{{\rm tr}}=n_0$. Afterward, the upper limit on $\tilde{\Lambda}$, from $\tilde{\Lambda}-q$ and $\Lambda_1-\Lambda_2$ diagrams for both events, imposed more stringent constraints on the speed of sound cases. Especially, the constraints from the first event (GW170817), which is more informative than the second one (GW190425), led to the additionally exclusion of EoSs with $n_{{\rm tr}}=1.5n_0$.

Subsequently, we introduced a new way to investigate and impose possible constraints on the transition density $n_{{\rm tr}}$  by using the upper limit on $\tilde{\Lambda}$ (derived from gravitational-wave observations) and taking advantage of the variety on $\tilde{\Lambda}$ for each EoS. In order to achieve that, we treated the transition density $n_{{\rm tr}}$ as a function of $\tilde{\Lambda}_{{\rm min,max}}^{(1/3,1)}$, where the parenthesis (min,max) denotes the minimum and maximum case respectively, for both sound speed bounds (indicated as (1,1/3)). From the first event (GW170817) we found that the speed of sound must be lower than the value  $v_s=c/\sqrt{3}$ at least up to densities $n_{{\rm tr}}\approx1.6 n_{\rm 0}$ and lower than $v_s=c$ up to densities $n_{{\rm tr}}\approx1.8n_0$. The respective values derived from the GW190425 event are $n_{{\rm tr}}\approx n_0$ for the lower speed of sound bound and $n_{{\rm tr}}\approx1.2n_0$ for the upper one. It is worth to point out that the event GW170817 offered more stringent constraints than the second one (GW190425). 

Moreover, we extended our approach by treating the effective tidal deformability $\tilde{\Lambda}$ as a function of the maximum mass $M_{\rm max}$ for both cases of speed of sound. From the first event (GW170817) we obtained that the maximum mass should be $M_{{\rm max}}\leq2.106\;M_\odot$ for the $v_s=c/\sqrt{3}$ bound and $M_{{\rm max}}\leq3.104\;M_\odot$  for the upper bound $v_s=c$. The limit of $M_{{\rm max}}\approx2.11\;M_\odot$ corresponds to a transition density equal to $n_{{\rm tr}}\approx1.5n_0$. According to this finding, the limit $v_s=c/\sqrt{3}$ is in contradiction with the observational evidence on the $M_{{\rm max}}$ of neutron stars and  must be violated in order to be able to simultaneously describe small values of the effective tidal deformability and high values for neutron star mass. This contradiction lays into the antagonized points of view; the upper limit on $\tilde{\Lambda}$ favors softer EoSs leading to higher values of $n_{{\rm tr}}$, while the observational information regarding to the maximum mass $M_{{\rm max}}$ of a neutron star requires stiffer EoSs, leading to smaller values of $n_{{\rm tr}}$. This contradiction blunts as the speed of sound takes larger values, leading to a maximum mass $M_{{\rm max}}\approx3.1\;M_\odot$ for the causal case $v_s=c$. We notice that the second event GW190425, was not able to offer further constraints.

Lastly, we studied the effective tidal deformability $\tilde{\Lambda}$ as a function of the radius $R_{1.4}$ of a $1.4\;M_\odot$ neutron star. All the EoSs follow a common trend, which is affected by the value of the chirp mass $\mathcal{M}_c$ of the binary system; higher values of $\mathcal{M}_c$ shift the trend downwards. From the event GW170817 we obtained an upper limit $R_{1.4}\approx13\;{\rm km}$ for both cases, which is consistent to other estimations. The event GW190425 provided an upper limit $R_{1.4}\approx14.712\;{\rm km}$ for the $v_s=c/\sqrt{3}$ bound and $R_{1.4}\approx14.53\;{\rm km}$ for the $v_s=c$ bound.

We postulate that future observations may offer even more rigorous constraints on the bound of the speed of sound. To be more specific, the detection of future events could provide further information on the upper limit of $\tilde{\Lambda}$, leading to more stringent constraints on $n_{{\rm tr}}$ and the sound speed bounds. According to our approach, the more informative events, for the lower limit of $n_{{\rm tr}}$, would be those with lighter masses. Also, it will be of great interest to impose constraints from the maximum value of the lowest limit of $\tilde{\Lambda}$. In this case, a lower limit derived from the EM counterpart of the events might be able to lead to the estimation of an upper value on the transition density $n_{{\rm tr}}$ where the two speed of sound bounds must be reached. We assume that despite the difficulties on the detection of the EM counterpart, the heavier neutron stars mergers could be more informative since they could impose an upper limit on $n_{{\rm tr}}$, than the lighter ones. Moreover, further detection of neutron stars mergers will assist both on the neutron stars maximum mass determination and its link to the speed of sound. Similarly, these detections will provide further information on the radius of neutron stars and it remains an open question its connection to the speed of sound bounds and the possible gain in their understanding.

%%%%%%%%%%%%%%%%%%%%%%%%%%%%%
\section*{Acknowledgments}
%%%%%%%%%%%%%%%%%%%%%%%%%%%%%%
The authors thank Assistant Prof. B. Farr and Dr. K. Chatziioannou for providing computational tools regarding the kernel density estimation of the sample data. The authors thank also Prof. K. Kokkotas for his constructive comments on the preparation of the manuscript.

%%%%%%%%%%%%%%%%%%%%%%%%%%%%%%%

%%%%%%%%%%%%%%%%%%%%%%%%%%%%%%

\end{document}